\PassOptionsToPackage{pdfborder={0 0 0},colorlinks=true, linkcolor=blue, urlcolor=blue, citecolor=blue}{hyperref}
\documentclass[lettersize,journal]{IEEEtran}
\usepackage{amsmath,amsfonts}
\usepackage{algorithmic}
\usepackage{algorithm}
\usepackage{array}
\usepackage[caption=false,font=normalsize,labelfont=sf,textfont=sf]{subfig}
\usepackage{textcomp}
\usepackage{adjustbox}
\usepackage{stfloats}
\usepackage{url}
\usepackage{orcidlink}
\usepackage{verbatim}
\usepackage{graphicx}
\usepackage{cite}
\usepackage{mdframed}

\newcommand{\updated}[1]{\textcolor{black}{#1}}

\usepackage{multirow} 
\usepackage{booktabs}
\usepackage{wrapfig}
\usepackage{enumitem}

\begin{document}

\title{Preserving Discrete Morse--Smale Complexes in \\ Error-Bounded Lossy Compression}

\author{Yuxiao~Li\,\orcidlink{0000-0002-8715-5982},
        Mingze~Xia\,\orcidlink{0009-0000-7653-9000},
        Xin~Liang\,\orcidlink{0000-0002-0630-1600},
        Bei~Wang\,\orcidlink{0000-0002-9240-0700},
        and~Hanqi~Guo\,\orcidlink{0000-0001-7776-1834}%
\thanks{Y. Li and H. Guo are with Ohio State University, Columbus, OH, USA. 
E-mails: \{li.14025, guo.2154\}@osu.edu}
\thanks{M. Xia, and X. Liang are with Oregon State University, Corvallis, OR, USA. 
E-mails: \{xiami, lianxin\}@oregonstate.edu}
\thanks{B. Wang is with University of Utah, Salt Lake City, UT, USA. 
E-mail: beiwang@sci.utah.edu}
}

\markboth{IEEE Transactions on Visualization and Computer Graphics,~Vol.~X, No.~X, X~2026}%
{Li \MakeLowercase{\textit{et al.}}: Preserving Discrete Morse--Smale Complexes in Error-Bounded Lossy Compression}

\maketitle
\begin{abstract}
Scientific applications are generating unprecedented volumes of data that overwhelm storage and transmission systems, posing significant challenges for the design of data management tools and scientific databases. Lossy compression has emerged as a promising strategy to address this problem, but most existing compressors fail to preserve the topology of scientific data, leading to inaccuracies in downstream analyses and potentially erroneous scientific conclusions.
In this work, we present a methodology for fully preserving the topology, specifically, Morse–Smale complexes (MSCs), in lossy-compressed 2D and 3D scalar field data from scientific simulations. We generalize the edit-based strategy introduced in MSz~\cite{li_msz} (a previous method that preserves only segmentations and cannot preserve saddles or separatrices) by extending the framework to the full MSCs, including all critical points and separatrices. Our approach corrects the MSCs in the decompressed output of any error-bounded lossy compressor (e.g., SZ3 or ZFP), referred to as the base compressor, using an iterative editing strategy that preserves all critical points and their connectivity via separatrices. During compression, we generate a sequence of quantized edits that are applied to the decompressed output, ensuring accurate preservation of topological features while maintaining the error within prescribed bounds. 
The strategy iteratively fixes critical points and separatrices in alternating steps until convergence is achieved in a finite number of iterations. To meet diverse application needs, our method offers flexible options (e.g., whether to preserve the geometry of separatrices) that balance compression efficiency with feature preservation. To reduce computation time, we leverage GPU parallelism to accelerate each component of the workflow. Experiments on multiple datasets demonstrate that our method achieves 100\% preservation of Morse–Smale complexes.
\end{abstract}


\section{Introduction}
\label{sec:intro}

The rapid advancement of scientific computing generates a large volume of data, such as in cosmology, combustion, and climate modeling, posing significant challenges to scientists in terms of data storage and visualization.
To this end, compression techniques, especially lossy compression that significantly reduces the size of scientific data, have been widely used in scientific data management and storage.  
In contrast to general-purpose compression, scientific applications often require preserving data fidelity within strict error bounds to ensure the validity of downstream analyses. As such, error-bounded lossy compression techniques have been widely used in scientific applications to address data challenges by achieving high compression ratios through strictly controlled errors in the decompressed data~\cite{sz, sz3, Tao_2017, Kai2021, Kai2020, zfp, FPZIP}.

\begin{figure}[htb!]
    \centering
    \includegraphics[width=\linewidth]{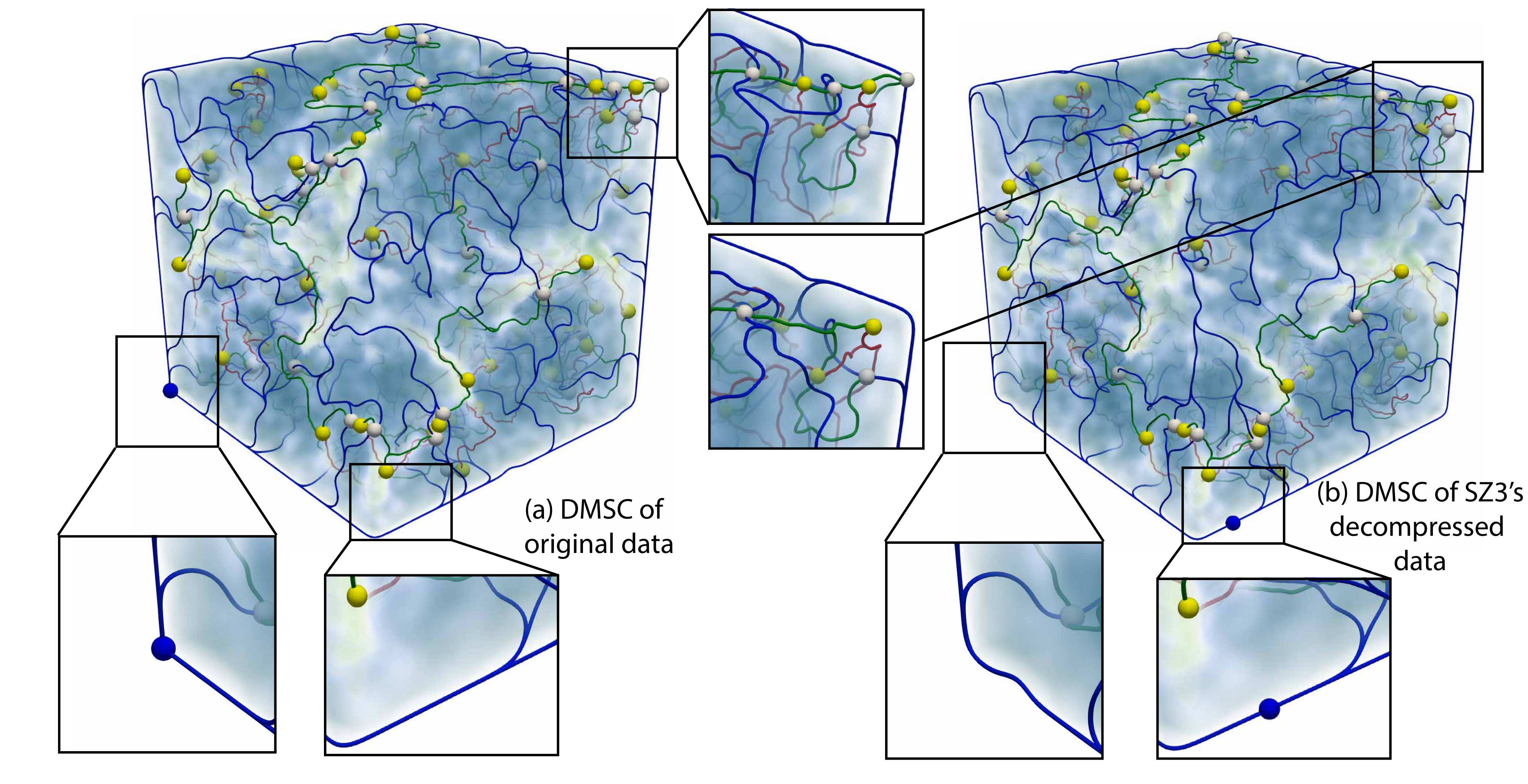}
    \caption{Morse--Smale complexes of the cosmology data with a topological simplification threshold~\cite{Edelsbrunner_persistence} of 0.2: (a) the original data and (b) SZ3's output (relative error bound = $1 \times 10^{-3}$).}
    \label{fig:cosmic}

\end{figure}

\begin{figure*}
  \includegraphics[width=\textwidth]{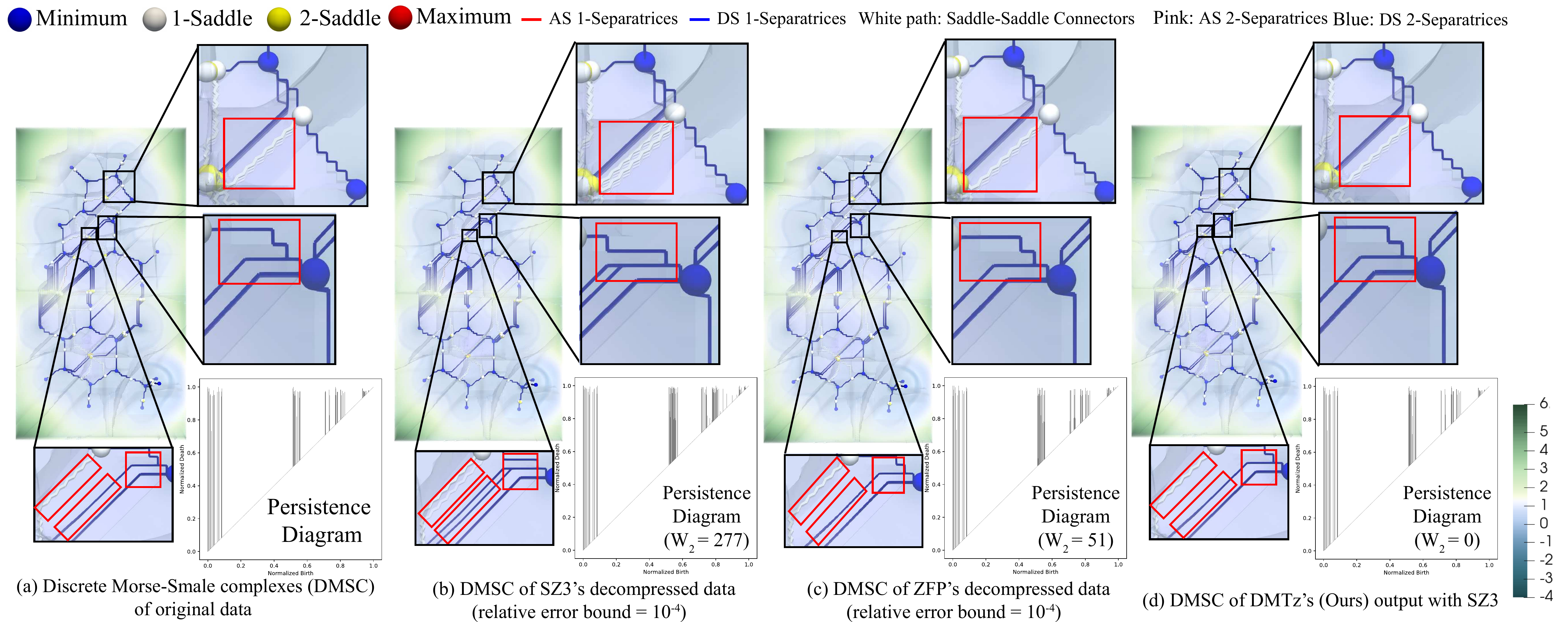}
  \caption{Morse--Smale complexes extracted from 3D scalar fields of a molecular dynamics simulation dataset Adenine Thymine (AT): (a) original data, (b) SZ3 decompressed data, (c) ZFP decompressed data, and (d) our method, obtained by modifying SZ3’s decompressed output to preserve the Morse--Smale complex. All compression results use a relative error bound of $10^{-4}$. Persistence diagrams derived from the Morse--Smale complexes are also shown for (a)-(d), with both birth and death values normalized. $W_2$ denotes the $L^2$ Wasserstein distance between each output’s persistence diagram and that of the original data, quantifying topological distortion.}
  \label{fig:teaser}
  
\end{figure*}

However, existing lossy compressors either do not consider the preservation of topology at all or only preserve part of the topology (e.g., critical points and topological segmentations)~\cite{yan2023toposz, Liang_2020, li_msz, gorski2025general, Xia24}, which impacts downstream analyses that rely on topological features. Compression-induced errors can distort the topology of the data, such as Morse--Smale complexes (MSCs)~\cite{msc, msc2}, as shown in Figure~\ref{fig:cosmic}, and their induced segmentations~\cite{MSS, li_msz}, and merge/contour trees~\cite{CARR200375, yan2023toposz,gorski2025general}, leading to inconsistencies between the topology of the original and decompressed data. Such topological inconsistencies may lead to the misinterpretation of scientific results and unreliable feature extraction in fields such as chemistry~\cite{Bhatia_2018, Günther_2014}, materials science~\cite{Petruzza_2020, Gyulassy_2007}, climate science~\cite {Dora_2013}, medical imaging~\cite{XU201438}, and cosmology~\cite{Shivashankar_2015}.

For example, in molecular electron density analysis~\cite{Bhatia_2018}, as shown in Figure~\ref{fig:teaser}(a), local maxima correspond to the positions of atomic nuclei, while local minima indicate low-density regions or internal cavities. Saddle points, where the density curves upward in one direction and downward in another, capture structures such as the centers of rings in molecules. The separatrices connecting these critical points represent atomic interaction paths, commonly interpreted as bond paths in chemical analyses. Thus, distortions introduced by lossy compression may significantly alter these critical features, potentially leading to severe errors in the interpretation of molecular structures. For instance, the disappearance of saddle points may disrupt separatrix structures, leading to the loss of topological representations of chemical bonds.

Topology inconsistencies introduced by compression are also a critical issue in time-dependent scientific simulations.
Although an individual timestep may be relatively small, large-scale studies typically contain hundreds or thousands of timesteps, each of which must be compressed and later analyzed.
Compression-induced distortions occur at every timestep, and the distortions accumulate across the sequence, degrading temporal analyses such as feature tracking, event detection, and persistence evolution, as demonstrated later in our experiments in Section~\ref{sec:time}.

While it seems possible to mitigate topological inconsistencies by storing the original topological structures alongside the compressed data, the stored structures still fail to align with the decompressed data. As demonstrated in Figure~\ref{fig:teaser}(b) and (c), separatrices extracted from the decompressed data are misaligned with the original structures, and storing the original topological structures no longer reflects the altered scalar field, resulting in inconsistencies in visualization and downstream analysis. For example, scientists may use the persistence diagram (a plot that tracks the lifespan of topological features) extracted from MSC to filter out less significant features, referred to as topological simplification~\cite{Edelsbrunner_persistence}. However, the persistence diagram extracted from the original MSC no longer aligns with the decompressed field, as shown in Figure~\ref{fig:teaser}. Therefore, rather than storing the original topological structures, it is essential to preserve the topology to ensure reliable scientific interpretation, as we demonstrated in our experiments in Appendix C.

\updated{This work targets the full preservation of MSC within decompressed data under error-bounded lossy compression.} The closest related method is MSz~\cite{li_msz}, which focuses on preserving MSC-induced segmentations, namely, piecewise linear MS segmentations (PLMSS). Note that preserving PLMSS does not preserve the full MSC; PLMSS offers a cheaper preview of MSC but lacks key topological constituents~\cite{MSS}, including saddles and saddle-related features. The lack of preservation of saddle-related features is a key limitation of MSz, as these features are fundamental to the structure of the MSC and influence a wide range of downstream analyses, such as persistence diagram computation and topological simplification~\cite{Edelsbrunner_persistence}, and topology-aware machine learning~\cite{pont2023}. In climate research, for instance, scientists rely on saddle points to identify potential turning points in storm paths~\cite{Dora_2013}. Distorted saddle points can mislead weather forecasts, potentially leading to inadequate responses to extreme weather events.

In this paper, we introduce \textbf{DMTz}, an iterative and edit-based algorithm for correcting MSCs in 2D/3D scalar field data that are compressed by arbitrary error-bounded lossy compressors. We generalize the edit-based paradigm in MSz~\cite{li_msz} by (1) introducing strategies for correcting full MSC instead of PLMSS and (2) improving the compression of edits by quantizing the corrections. DMTz also strictly controls the pointwise error between the decompressed and original data, consistent with the bounded local error guarantees supported by most scientific data compressors for other tasks such as checkpoint-restarting~\cite{SHAHZAD2013}. Specifically, we target the widely used MSC computation scheme based on discrete Morse theory, referred to as the discrete Morse--Smale complex (DMSC), because of its robust implementation and adoption in topological visualization tools such as the Topology Toolkit (TTK)~\cite{masood2021overview}. For a comprehensive comparison between DMTz and MSz, see Section~\ref{sec:msz_cp}.

DMTz uses an iterative workflow with two subloops: (1) the critical cells loops (C-loops), which preserve the type and location of the critical cells (equivalent to critical points in discrete Morse theory), and (2) the separatrices loops (S-loops), which detect and fix the inaccuracies in the separatrices. Through a finite number of iterations, the strategy identifies a subset of data that requires edits, ensuring that the topology in the decompressed data remains consistent with that in the original dataset while guaranteeing the global error bound. Moreover, we enhanced the edit compression method by converting most edits into a quantized form, reducing the average storage overhead for storing the edits. We also used GPU parallelism to accelerate each component.

We further propose a multitier topology preservation paradigm to support diverse scientific needs, enabling users to balance compression efficiency and topological preservation. For example, medical imaging applications may focus on preserving extrema, such as intensity peaks corresponding to tumors~\cite{XU201438}, and atmospheric river analysis often relies on preserving the connectivity structure or topological skeleton~\cite{Lan_ivt}, while combustion simulations require accurate preservation of saddle points and their connections to extrema~\cite{Bremer_combustion, Bremer_flames}. 

In summary, the contributions of this paper are:
\begin{itemize}[noitemsep,leftmargin=*]
    \item We develop an iterative strategy for preserving Morse--Smale complexes from lossy-compressed 2D/3D scalar fields, which is theoretically applicable to any existing error-bounded lossy compressor.
    \item We generalize the method for compressing edits in MSz~\cite{li_msz} by converting most edits into a quantized form, thereby reducing overall storage overhead.
    \item We conduct a comprehensive evaluation on diverse datasets from multiple applications using four off-the-shelf base compressors: SZ3~\cite{sz3}, ZFP~\cite{zfp}, SPERR~\cite{SPERR}, and MGARD~\cite{MGARD}.
\end{itemize}

\section{Related Work}
We review the related work on lossy compression and topology-preserving compression.
\subsection{Lossy Compression for Scientific Data}
Lossy compression methods are categorized as error-bounded and non-error-bounded based on whether pointwise error is limited by user-defined bounds. Non-error-bounded methods often achieve higher compression ratios while not constraining pointwise error within user-defined bounds. For example, neural network-based approaches, such as autoencoders~\cite{Liu_AE} and implicit neural representations~\cite{Neurcomp, NeRVI}, optimize global reconstruction quality without error guarantees. We focus on error-bounded lossy compression that offers more precise control over data distortion, as discussed below.

Error-bounded lossy compression achieves efficient data compression while ensuring that the introduced error remains within the user-defined error bound, providing high data quality. Error-bounded compression methods can be divided into prediction-based and transformation-based approaches. Prediction-based methods, like the SZ series~\cite{sz, sz3, Tao_2017, Kai2021,Qoz, huang2024cuszp2, huang2023cuszp, huang2025lscomp}, estimate data points using predictors such as Lorenzo and then quantize the residuals for compression. Recent works further explore neural networks to improve prediction accuracy, including AE-SZ~\cite{liu_2021} and SRNN-SZ~\cite{liu2023srnsz}. FPZIP~\cite{FPZIP} and ISABELA~\cite{ISABELA} follow similar prediction-based approaches with bit-plane truncation and B-spline transformations, respectively. Transform-based compressors, like ZFP~\cite{zfp}, TTHRESH~\cite{TTHRESH}, SPERR~\cite{SPERR}, and MGARD~\cite{MGARD}, apply techniques like wavelet transforms or tensor decompositions to compress data more efficiently. For a more comprehensive survey of error-bounded lossy compression methods for scientific datasets, we refer readers to Di et al.~\cite{di2024survey}.

\subsection{Topology-Preserving Compression of Scalar-Field Data}\label{sec:topo_compression}
Prior work has studied topology-preserving compression for various topological descriptors~\cite{yan2023toposz, gorski2025general, Liang_2020, LiangDCRLOCPG23, Xia24, Xia25, li2026pmsz}. \updated{Our work focuses on the preservation of the Morse--Smale complex (MSC) under error-bounded lossy compression with strict pointwise error guarantees.} Note that the pointwise error bound is still important in the context of topology-preserving compression beyond topological data analysis (TDA), as many downstream applications and checkpoint-restart simulations rely on local data accuracy, and uncontrolled errors can significantly affect the integrity of scientific results, such as statistical analysis. To achieve topological preservation in error-bounded lossy compression, one needs to modify the compression workflow. We classify these modifications into the following three categories.

The first strategy is to \textbf{modify the input data} before compression to guide the preservation of topological features. For example, Soler et al.~\cite{soler2018topologically} proposed a topology-controlled compression method that preserves the persistence diagram by adaptively quantizing data based on a persistence simplification threshold. Their approach relies on the input of the persistence threshold for pointwise error control. In contrast, our method targets a different topological descriptor, MSC, and is parameter-independent.

The second strategy is to \textbf{modify compression algorithms} directly, but the modification is specific to a compressor, limiting the broader applicability. For example, Yan et al.~\cite{yan2023toposz} proposed TopoSZ, using topological constraints derived from segmentations guided by contour trees by modifying the SZ~\cite{sz} compression algorithm. 

The third strategy is to \textbf{correct the decompressed data} to preserve topological features, but existing methods do not support the preservation of the full MSC. For example, Gorski et al.~\cite{gorski2025general} proposed a framework that clamps the values of decompressed data to preserve contour trees. Li et al.~\cite{li_msz} proposed MSz, an edit-based method for preserving MS segmentations in 2D/3D piecewise linear scalar fields by focusing on extrema and the integral lines connecting them by deriving a series of edits applied to the decompressed data during compression. 

As related, the HPC and visualization communities have also explored topology preservation in vector fields. For example, Liang et al.~\cite{Liang_2020, LiangDCRLOCPG23} and Xia et al.~\cite{Xia24, Xia25} proposed compressor-specific strategies that preserve critical points and topological skeletons by incorporating topological constraints into the compression pipeline. Theisel et al.~\cite{theisel2003combining} proposed a method that first modifies the vector field to simplify its topology before applying compression. Tricoche et al.~\cite{tricoche2001continuous} collapse edges in a 2D mesh to guarantee topology preservation.

\subsection{Topological Simplification and Smoothing}
\updated{Topological simplification and smoothing are closely related to topology-aware processing and establish consistency between modified data and topological descriptors, forming a conceptual foundation closely related to topology-preserving compression. As general techniques, they are widely used in data analysis and visualization to reduce topological complexity or remove noise by modifying the data. However, these approaches are typically not designed to retain detailed topological structures under strict pointwise error bounds. Our work, in contrast, addresses a different problem setting, where both topology preservation and strict error bounds must be satisfied simultaneously under error-bounded lossy compression.}

\updated{Topological simplification and smoothing techniques have been studied for both scalar fields and vector fields.
For scalar fields, simplification techniques remove low-persistence features by canceling pairs of critical points~\cite{Gyulassy_06, Tierny_12}. Optimization-based methods, such as solver-based smoothing~\cite{Kissi_25}, adjust scalar values to approximate a target persistence diagram. Other techniques reconstruct the scalar field using monotonicity constraints~\cite{Gunther_14}, enforce smoothness via $C^1$-continuity while preserving selected features~\cite{Weinkauf_10}, or generate smooth shape-aware functions with controlled extrema~\cite{jacobson2012smooth}.}

\updated{For vector fields, simplification methods focus on reducing topological noise while retaining important flow features. Early foundational works by Tricoche et al.~\cite{tricoche2000topology, tricoche2001continuous} introduced continuous topology simplification for planar vector fields, while Theisel~\cite{theisel2002designing} explored designing 2D vector fields with arbitrary topologies. Building upon these principles, Weinkauf et al.~\cite{Weinkauf_05} extract higher-order critical points in 3D fields, and Wang et al.~\cite{Wang_Robustness} introduce hierarchical simplification based on robustness for steady and unsteady 2D fields.}

\section{Background}
We review the background of Morse--Smale complexes and topology preservation with edits in error-bounded lossy compression. 

\subsection{Morse--Smale Complexes}
MSC is one of the most widely used topological descriptors for capturing the structure of scalar fields. 
By identifying critical points and their connectivities via integral curves, MSC decomposes the domain into regions of consistent gradient behavior (i.e., by following the gradient line, each data point in the same region reaches the same maximum or minimum), revealing important topological features in the data.
Two distinct theoretical frameworks exist for computing MSC: \emph{discrete Morse theory}~\cite{FORMAN199890} and \emph{piecewise linear (PL) Morse theory}~\cite{cp,msc2}. Discrete Morse theory has gained broad adoption because of its robustness and serves as the foundation for widely used topological analysis libraries such as the Topology ToolKit (TTK)~\cite{masood2021overview}; for a detailed comparison between the discrete and PL Morse theory, refer to~\cite{De_2015}.

\begin{figure}[htb!]
    \centering
    \includegraphics[width=\linewidth]{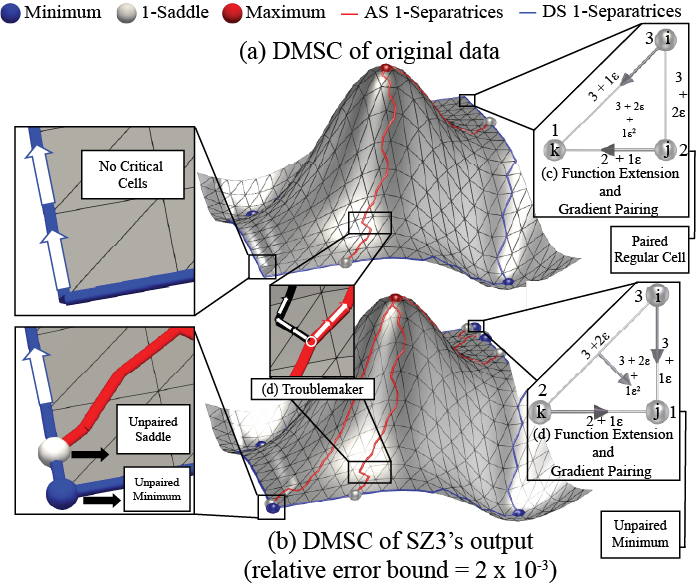}
    \caption{Impact of SZ3 on the discrete Morse--Smale complex (DMSC) for synthetic data under a relative error bound of $2 \times 10^{-3}$. DMSCs of (a) the original data and (b) SZ3’s output are shown, while (c) and (d) illustrate the processes of function extension and gradient pairing, respectively.}
    \label{fig:synthetic}
    
\end{figure}

\textbf{Discrete Morse theory} introduced by Forman~\cite{Forman_dmt}, defines a discrete version of gradient field (namely, \emph{discrete gradient field}) and integral lines (namely, V-paths) on cell complexes. Formally, a (simplicial) \emph{cell complex} is an abstract representation of a mesh composed of \emph{0-cells} (vertices), \emph{1-cells} (edges), \emph{2-cells} (triangles), and higher-dimensional cells. A \emph{facet} refers to the $(k-1)$-dimensional face of a k-dimensional cell, while a \emph{cofacet} is the $(k+1)$-dimensional cell that contains the k-dimensional cell as its face. A \emph{discrete gradient field} consists of pairings between incident cells whose dimensions differ by one, with the constraint that each cell appears in at most one pair. For example, in Figure~\ref{fig:synthetic}(c), vertex $i$ is paired with its cofacet edge $ij$ (the rationale for such pairings will be explained later). Cells that are not involved in any pairing are identified as \emph{critical cells}, which correspond to critical points in the data. For instance, vertex $j$ in Figure~\ref{fig:synthetic}(d) is a minimum because it is unpaired. A \emph{gradient path} is a sequence of cells with alternating dimensions paired through the gradient pairing process (e.g., a path with 0-, 1-, 0-, …, 0-cells, such as the blue curve in Figure~\ref{fig:synthetic}), and gradient paths that connect critical cells correspond to separatrices in Morse theory, as the red and blue paths shown in Figure~\ref{fig:synthetic} (a) and (b). Separatrices that connect 1-saddles to 2-saddles are specifically referred to as saddle-saddle connectors.

\textbf{Function extension.} A critical step in computing MSC from a scalar field is extending scalar values from vertices to higher-dimensional cells to enable valid gradient pairings, as scalar fields in practice are typically defined only on vertices. There exist several function extension methods for constructing discrete Morse functions, and we adopt the method of Shivashankar et al.~\cite{SN12} due to its compatibility with dimension-wise parallelism, which enables independent processing of vertices, edges, triangles, and higher-dimensional cells and aligns well with our staged edit-based strategy.
Specifically, the discrete Morse function $F$ is recursively extended to a $d$-dimensional cell $\alpha$ as:
\begin{equation}\label{equation:discrete_value}
    F(\alpha) = F(G_0(\alpha)) + \epsilon^d F(G_1(\alpha)),
\end{equation}
where $\epsilon$ is a (symbolic) perturbation to enforce strict ordering, $G_0(\alpha)$ is the highest-valued face of $\alpha$, and $G_1(\alpha)$ is the highest among the remaining non-adjacent faces. We define $P_{\alpha}$ as the set of $(d+1)$-dimensional cofacets $\beta$ that $\alpha$ is the highest-valued facet of $\beta$ under the extended function. For example, for edge $ij$ in Figure~\ref{fig:synthetic} (c), $G_0(ij)$ is vertex $i$, $G_1(ij)$ is vertex $j$ and $P_{ij}$ is triangle $ijk$.

\textbf{Gradient pairing and integral line tracing.} Figure~\ref{fig:synthetic}(c) illustrates the function extension and gradient pairing schemes used to construct the discrete gradient vector field. The scalar field $f$ is defined only on vertices: $f_i = 3$, $f_j = 2$, and $f_k = 1$. These values are recursively extended to higher-dimensional cells. For example, the extended function value of edge $ij$ is $F_{ij} = f_i + \epsilon f_j = 3 + 2\epsilon$, and for edge $ik$, $F_{ik} = 3 + \epsilon$.

Based on the extended function $F$ of the original scalar field $f$, the discrete gradient field is constructed by pairing each $d$-dimensional cell $\alpha$ with the $(d{+}1)$-dimensional cofacet in $P_{\alpha}$ that has the smallest function value. For example, in Figure~\ref{fig:synthetic}(c), vertex $i$ is the highest-valued facet of both edges $ij$ and $ik$. Because $F_{ik} = 3 + \epsilon < F_{ij} = f_j = 3 + 2\epsilon$, edge $ik$ is the lowest-valued cofacet in $P_i$, and vertex $i$ is therefore paired with edge $ik$.
 
The function extension and gradient pairing method implies that errors introduced by lossy compression can propagate from vertex scalar values to the extended function values of higher-dimensional cells, leading to incorrect gradient pairings and distorting the topology.
As illustrated in Figure~\ref{fig:synthetic}(c), in the original data $f$, the pairing of vertex $j$ is paired with edge $jk$ because $ f_j = 2 > f_k = 1$. However, in the decompressed data $\hat{f}$ (Figure~\ref{fig:synthetic}(d)), compression-induced error alters the vertex scalar values such that $\hat{f}_j = 1 < \hat{f}_k = 2 < \hat{f}_i = 3$. As a result, vertex $j$ (originally a regular cell paired with edge $jk$) becomes unpaired and is incorrectly identified as a minimum.

\subsection{Edit-based Paradigm for Correcting Topology in Lossy Compressed Data}\label{sec:edit_method}

This paper adopts and generalizes the edit-based strategy introduced in MSz~\cite{li_msz}, which generates a series of targeted edits to the decompressed data to recover the topological structures while maintaining the prescribed error bounds~$\xi$. This work extends the edit-based strategy in two key ways: (1) targeting the preservation of a full discrete Morse--Smale complex (DMSC) instead of a piecewise linear Morse--Smale segmentation (PLMSS), and (2) encoding most of the edits in quantized form to reduce storage overhead.
\begin{wrapfigure}{R}
    {1.5in}
    \vspace{-0.15in}
    \includegraphics[width=1\linewidth]{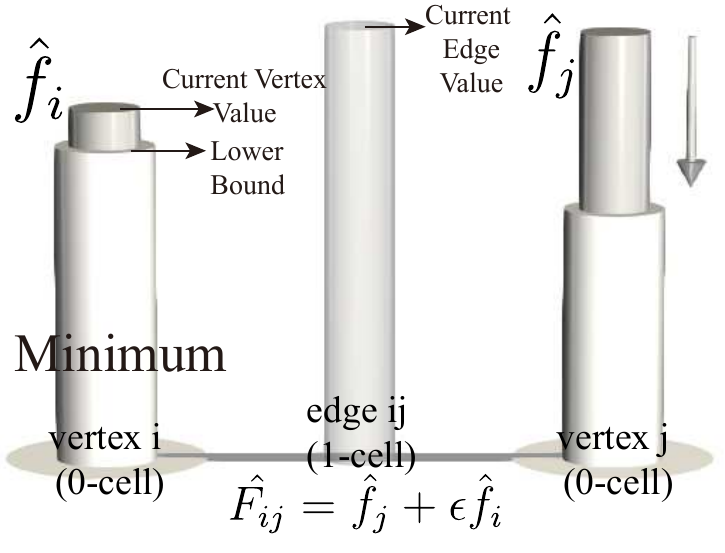}
    \caption{Illustration of the edit-based strategy: vertex $i$, originally a regular cell paired with edge $ij$, becomes an unpaired minimum after decompression. The height of each bar above a cell represents its current scalar value.}
    \label{fig:edits} 

\end{wrapfigure}

The edit-based strategy is effective for preserving feature descriptors relying on the ordering of data, such as MS segmentations and MSC, because it modifies the values of a subset of data points to preserve the local data ordering necessary for the target features. Because the edits are directly applied to the decompressed data, this method can be applied to any existing error-bounded lossy compressor without being tied to the specific compression algorithm. 

The edit-based strategy involves identifying distorted features, such as critical points, during each iteration and modifying the scalar value of specific data points to correct distortions. Assume that vertex $i$ is a regular cell paired with edge $ij$ in the original data $f$ and vertex $j$ is one of its neighbors, such that $f_i > f_j$. In the decompressed data $\hat{f}$ as shown in Figure~\ref{fig:edits}, vertex $i$ becomes a minimum as $\hat{f_i} < \hat{f_j}$, we can fix the topological inconsistency by decreasing the value of the vertex $j$ such that $\hat{f_i} > \hat{f_j}$, making vertex $i$ paired with edge $ij$. However, the modification may introduce new distortions, leading to further iterations. The process continues until all targeted features are preserved.

The key to the convergence of the iterative process is that the edits applied to each data point are either negative or zero. Specifically, for a data point $i$ at iteration $k$, the edited scalar value $g_i^{(k)}$ satisfies: $\hat{f}_i = g_i^{(0)} \geq \cdots \geq g_i^{(k)} \geq g_i^{(k+1)} \geq \cdots \geq f_i-\xi$, where $\hat{f}_i$ is the initial decompressed value, $f_i$ is the original value, and $\xi$ is the absolute error bound. Since the original data has an inherent order, and the edits on data point $i$ are negative or zero, one can always find a finite iteration $k$ where the scalar value of $i$ and its neighbors align with the original data’s order. In the worst-case scenario, all data values may decrease to the lower bound $f_i - \xi$, under which the data ordering will remain consistent with that of the original data, thus guaranteeing convergence. 

While edit-based methods enable topological feature preservation, storing edits comes with a cost. To address the storage overhead introduced by edits, we propose a quantization strategy (referred to as \emph{quantized edits}) that stores most edits in a quantized form, with only a small subset needing to be stored losslessly. The quantized edits reduce the overall storage cost without affecting feature accuracy, in contrast to MSz, which stores all edits losslessly (referred to as \emph{floating-point edits}). Later, we demonstrate the effectiveness of our quantized edits in our experiments.

\section{Our Method}
In this section, we formally define the problem of preserving the DMSC in error-bounded lossy compression and present our methodology, which iteratively corrects compression-induced distortions while ensuring the error bound is maintained.

Notations used throughout our paper are as follows: $\xi$ represents the user-defined error bound, $i$ denotes the $i$th vertex, edge $ij$ refers to the edge consisting of vertices $i$ and $j$, triangle $ijk$ denotes the triangle formed by vertices $i$, $j$ and $k$, and tetrahedron $ijkl$ represents the tetrahedron consisting of vertices $i$, $j$,
$k$ and $l$. The scalar values at vertex $i$ in the original and decompressed data are represented by $f_i$ and $\hat{f}_i$, respectively, while $g_i$ refers to the edited scalar value at vertex $i$. For a given cell $\alpha$, $P_{\alpha}$ denotes the set of $(d+1)$-dimensional cofacets $\beta$ for which $\alpha$ is the maximum-valued facet of $\beta$.
\begin{figure*}
    \centering
    \includegraphics[width=\textwidth]{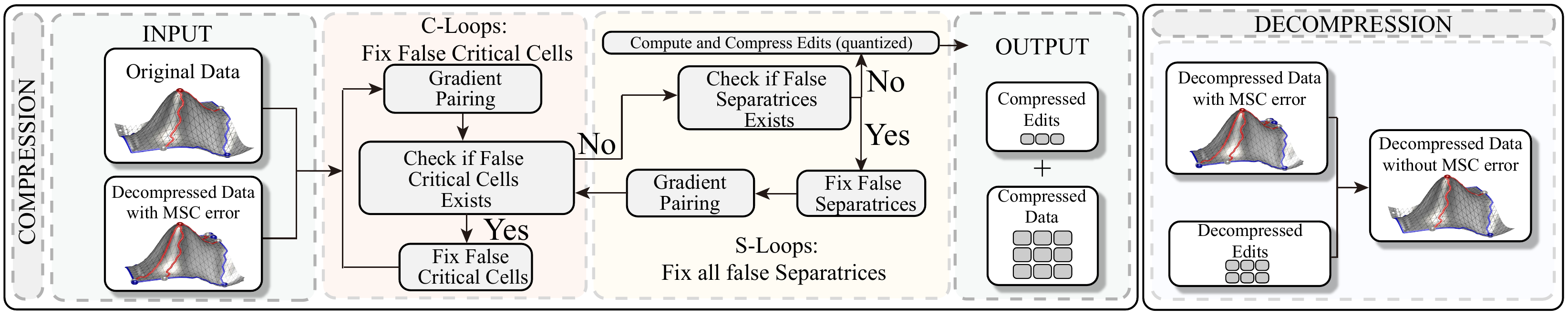}
    \caption{Workflow of our method for preserving the full Morse--Smale complex (MSC). During compression, the algorithm generates a sequence of edits through two alternating loops: (1) C-loops, which iteratively correct false critical cells, and (2) S-loops, which iteratively correct false separatrices. These loops alternate until no false critical cells or separatrices remain. The resulting quantized edits are losslessly compressed and stored alongside the compressed data. During decompression, the edits are applied to the reconstructed data to restore the MSC.}
    \label{fig:workflow}
    \vspace{-1em}
\end{figure*}
\subsection{Problem Statement}

We formulate the preservation of MSC in 2D and 3D scalar fields under error-bounded lossy compression. The inputs of our algorithm include the original scalar field $f$ and the decompressed field $\hat{f}$, both defined on the same cell complex with the same number of vertices, where the underlying cell complex and vertex connectivity remain unchanged. 
The output of our algorithm is a set of edits $\{\delta_i\}$, where each edit adjusts the scalar value of a vertex $i$ to obtain the final edited value $g_i = \hat{f}_i + \delta_i$.  With the edits, our method still ensures that the final edited scalar field $g$ strictly satisfies the user-prescribed absolute error bound $\xi$, that is, $|f_i - g_i| \leq \xi$; more importantly, the edited data have the identical topology as the original data. 

\textbf{Degeneracy handling (e.g., flat plateaus).} As a fine detail in our problem formulation, we assume all scalar fields are \emph{Morse functions}, i.e., for any two vertices $i$ and $j$, $f_i \neq f_j$.  In practice, to handle degenerate cases in non-Morse functions, we apply a commonly used technique in computational topology, namely simulation of simplicity (SoS)~\cite{Edelsbrunner_persistence} that introduces a symbolic perturbation. Symbolic perturbations address equal function values by introducing a consistent symbolic ordering that distinguishes between them; for example, when two vertices $i$ and $j$ have equal scalar values, we enforce $f_i > f_j$ if $i > j$.  As such, even in extreme cases where function values plateau, our method is robust and preserves the topology as long as the MSC computation uses the same symbolic perturbation schemes, as discussed in Appendix B.

\textbf{Multitier preservation targets.} 
Driven by diverse application needs, we further define multitier preservation targets to capture varying levels of MSC fidelity, enabling users to balance compression performance and structural accuracy. 
In Tier 1 (Extrema Preservation), the focus is on preserving the maxima and minima. In this case, any cell identified as an extremum in the original data retains both its critical type and its location in the decompressed data, whereas cells that are not extrema in the original remain non-extrema after decompression.
Building upon this, Tier 2 (Critical Cells Preservation) extends Tier 1 by additionally preserving saddle points.
Further, Tier 3 (Connectivity Preservation) guarantees that each saddle in the decompressed data connects to the same set of extrema as in the original, while still allowing the geometric shapes of separatrices to be distorted.
Finally, in Tier 4 (Separatrix Preservation), the separatrices that connect critical cells are also preserved, ensuring that they follow the same paths in the decompressed data as in the original.

\subsection{Method Overview}

Our methodology corrects distortions in DMSC by editing vertex values to fix incorrect gradient pairings introduced by compression. We design an iterative workflow to compute vertex-wise edits during compression, as shown in Figure~\ref{fig:workflow}. The workflow alternates between two main loops: \textbf{critical cells loops} (C-loops): correct false critical cells, preserving extrema and saddles (T1, T2–T4). \textbf{separatrices loops (S-loops):} correct false separatrices, preserving saddle-extrema connectivity (T3) and geometric separatrices (T4). 

The key insight is that modifying vertex values directly alters the extended function values of higher-dimensional cells (edges, triangles, tetrahedra) through Equation~\eqref{equation:discrete_value}. Since gradient pairings depend entirely on the extended function values, adjusting vertex values enables us to recover the correct pairings and preserve DMSC. For a better understanding of our methodology, we summarize the pairing rules derived from the approach proposed by Shivashankar et al.~\cite{SN12} after function extension by using Equation~\eqref{equation:discrete_value}: 
\textbf{vertex $i$} is paired if there exists at least one adjacent vertex $j$ such that $f_i > f_j$, and it pairs with the edge $ij$ where $f_j$ is minimal among all such $j$; 
\textbf{edge $ij$} is paired if either (1) it has been paired by a vertex, or (2) there exists a triangle $ijk$ such that $\min(f_i, f_j) > f_k$, in this case, the edge pairs with the triangle $ijk$ where $f_k$ is minimal among all such $k$; 
\textbf{triangle $ijk$} is paired if either (1) it is paired with an edge, or (2) there exists a \textbf{tetrahedron} $ijkl$ such that $\min(f_i, f_j, f_k) > f_l$, in this case, the triangle pairs with the tetrahedron $ijkl$ where $f_l$ is minimal among all such $l$.

Specifically, in each loop, we identify every vertex $i$ that causes incorrect gradient pairing. This vertex is either part of a false critical cell or a \emph{troublemaker} (the first cell along a separatrix where the gradient pairing result is incorrect, as shown in Figure~\ref{fig:synthetic}(d)). Once identified, we gradually decrease its scalar value to recover the correct gradient pairing result. Let $q_{\max}$ be a user-defined hyperparameter that controls the edit step size, where each edit decreases the scalar value of a vertex $i$ by $\xi/2^{q_{\max}}$, $q$ denote the number of edits applied to $i$, and $g_i^{(k)}$ is the edited value at iteration $k$. The scalar value of vertex $i$ is updated iteratively as follows: 

\begin{equation}\label{eq1}
\begin{adjustbox}{max width=\linewidth}
$
g_{i}^{(k+1)} =
\begin{cases}
g_{i}^{(k)} - \xi / 2^{q_{\max}}, & \text{if } q < q_{\max} \text{ and } g_{i}^{(k)} - \xi / 2^{q_{\max}} \ge f_i - \xi, \\
f_i - \xi, & \text{otherwise}.
\end{cases}
$
\end{adjustbox}
\end{equation}

The edit continues until either $q$ reaches $q_{\max}$ or further editing exceeds the lower bound $f_i - \xi$. In either case, the final edit $\delta_i$ is stored as a floating-point number instead of a quantized number. For example, assume that vertex $i$ is first edited from $\hat{f}_i$ to $g_i^{(1)} = \hat{f}_i - \xi/2^{q_{\max}}$, then to $g_i^{(2)} = \hat{f}_i - 2 \cdot \xi/2^{q_{\max}}$, and so on. The edit continues until either $q = q_{\max}$, or $g_i^{(k+1)} \leq f_i - \xi$. At that point, we set $g_i = f_i - \xi$, and losslessly store the edit $\delta_i = f_i - \xi - \hat{f}_i$.

\subsection{Critical Points Loops (C-Loops)}\label{sec:cloop}
The C-loops focus on fixing all false critical cells, including false positive/negative minima, saddles, and maxima. \emph{False positive} cases are defined as cells paired with their facet/cofacet in the original data but not paired with any in the decompressed data. Conversely, \emph{false negative} cases are cells that were originally unpaired but paired with one of their facets/cofacets in the decompressed data.
We will discuss the method for selecting the vertex $i$ that needs editing in different cases below. For false positive/negative minima, we describe the fix process in detail, while for the remaining cases, we focus on the overall editing strategy and omit intermediate steps for brevity.

\begin{figure}[htb!]
    \centering
    \includegraphics[width=0.8\linewidth]{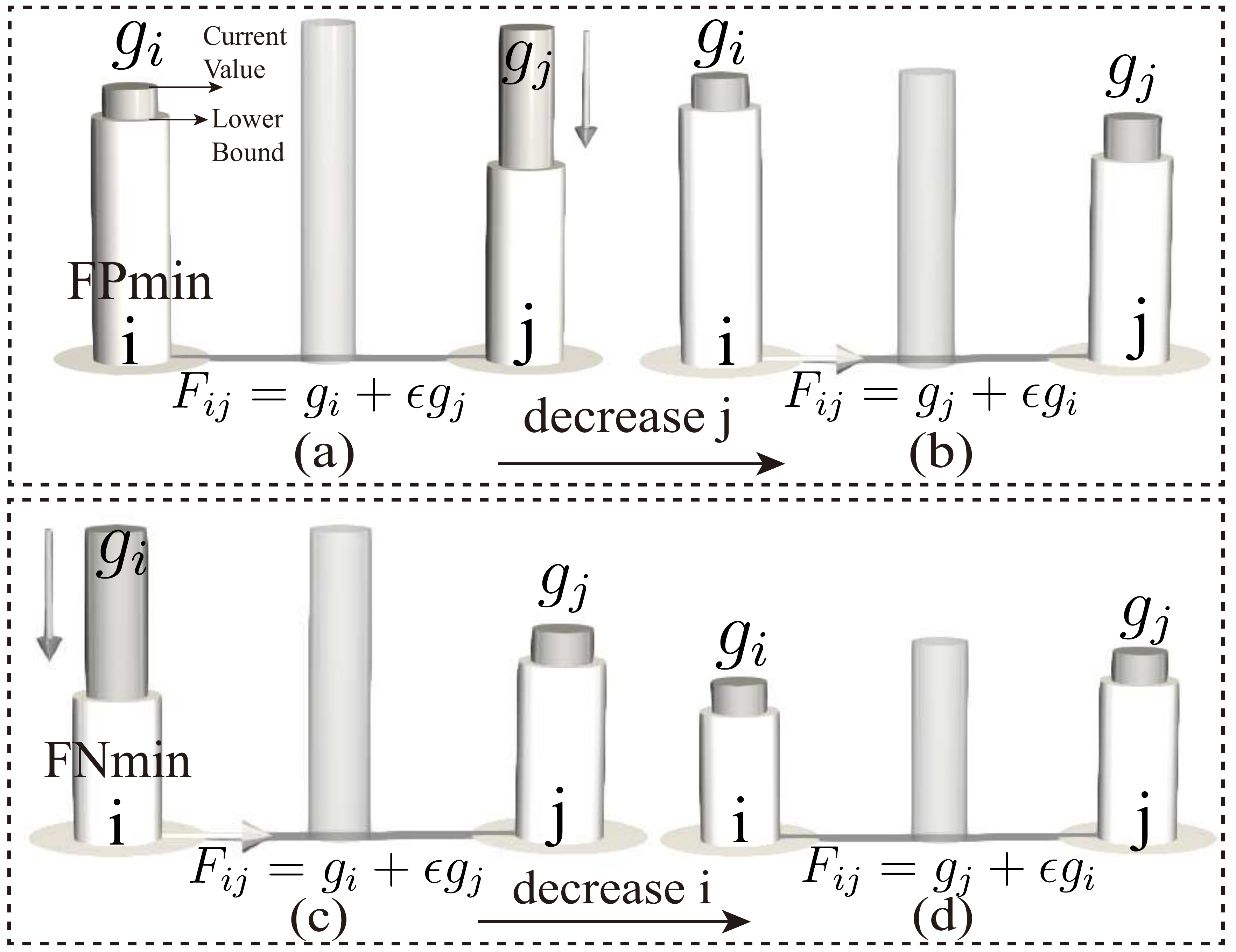}
    \vspace{-2mm}
    \caption{Fixing an FPmin/FNmin vertex $i$. Panels (a) and (b) illustrate the correction of an FPmin, while (c) and (d) show the correction of an FNmin. The height of the white cylinder above each vertex represents its lower bound ($f - \xi$), the gray cylinder indicates its current value, and the translucent gray cylinder above each edge denotes the current extended function value. Arrows from vertices to edges indicate the resulting gradient pairings.} 
    \label{fig:min}
\end{figure}

\noindent \textbf{False Positive Minimum (FPmin).} 
In the original data, a regular vertex $i$ is paired with an adjacent edge $ij$ if its scalar value is greater than that of its neighbor $j$ (i.e., $f_i > f_j$). A False Positive Minimum (FPmin) occurs when compression inverts this local gradient. Specifically, after decompression, if $\hat{f}_i$ becomes less than all its neighbors ($\hat{f}_i < \hat{f}_k$ for all adjacent neighbor $k$ of $i$), vertex $i$ fails to pair with any edge and is incorrectly classified as a minimum.

We use Figure~\ref{fig:min} to explain this case further. Assume vertex $i$ is originally paired with edge $ij$ ($f_i > f_j$). However, after compression, as shown in Figure~\ref{fig:min}(a), vertex $i$ becomes an FPmin because $\hat{f}_i < \hat{f}_j$. As a result, vertex $i$ remains unpaired and is incorrectly identified as a minimum. To correct this, our method identifies vertex $j$ for editing. By decreasing $\hat{f}_j$ as shown in Figure~\ref{fig:min}(b), the condition $g_i > g_j$ is restored. This enables vertex $i$ to pair with edge $ij$, eliminating the FPmin.

\noindent \textbf{False Negative Minimum (FNmin).} 
Assume that vertex $i$ is an unpaired minimum in the original data, meaning that $f_i < f_j$ for all adjacent vertices j. After decompression, as illustrated in Figure~\ref{fig:min}(c), $\hat{f}_i> \hat{f}_j$, making vertex $i$ incorrectly paired with edge $ij$, turning it into an FNmin. To fix this, we decrease $g_i$ until $g_i < g_j$. Once corrected, vertex $i$ no longer dominates any adjacent edge and returns to being an unpaired minimum, as shown in Figure~\ref{fig:min}(d).

\begin{figure}[htb!]
    \centering
    \includegraphics[width=\linewidth]{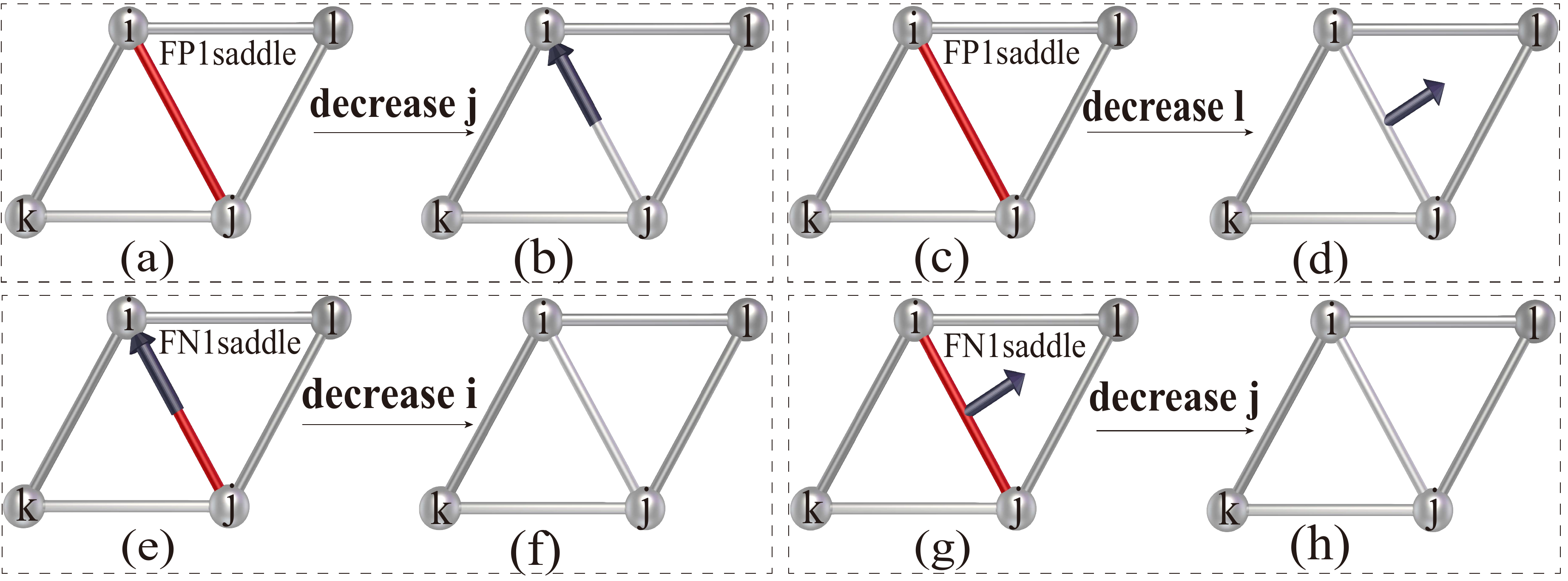}
    \caption{Fixing an FP1saddle that should be paired with a vertex, as shown in (a) and (b), or with a triangle, as shown in (c) and (d); and an FN1saddle that is incorrectly paired with a vertex, as shown in (e) and (f), or with a triangle, as shown in (g) and (h).}
    \label{fig:saddle}
    
\end{figure}

\noindent \textbf{False Positive 1-Saddle (FP1saddle).} 
According to the gradient pairing process, an FP1saddle edge $ij$ in the original data should be paired in one of the two conditions: \textbf{1}. paired with a vertex $ v_i $, requiring $f_i > f_j$; \textbf{2}. paired with a triangle $ijl$, requiring $ f_i > f_l $ and $ f_j > f_l $. However, in the decompressed data, the vertices forming the FP1saddle edge $ij$ do not satisfy either of the above conditions; therefore, we can modify the gradient pairing result by editing one of the vertices associated with the vertex or triangle paired with edge $ij$ in the original data, as shown in Figure~\ref{fig:saddle} (a)-(d).

\noindent \textbf{False Negative 1-Saddle (FN1saddle)} could be fixed by making it unpaired in the decompressed data. If an FN1saddle edge $ij$ is paired with vertex $i$ in the decompressed data (as shown in Figure~\ref{fig:saddle}(e) and (f)), we decrease vertex $i$ so that $g_i < g_j $, preventing edge $ij$ from pairing with vertex $i$. Similarly, if it is paired with triangle $ijl$ (as shown in Figure~\ref{fig:saddle}(g) and (h)), we decrease either vertex $i$ or vertex $j$, ensuring that edge $ij$ cannot pair with triangle $ijl$. 

\begin{figure}[htb!]
    \centering
    \includegraphics[width=\linewidth]{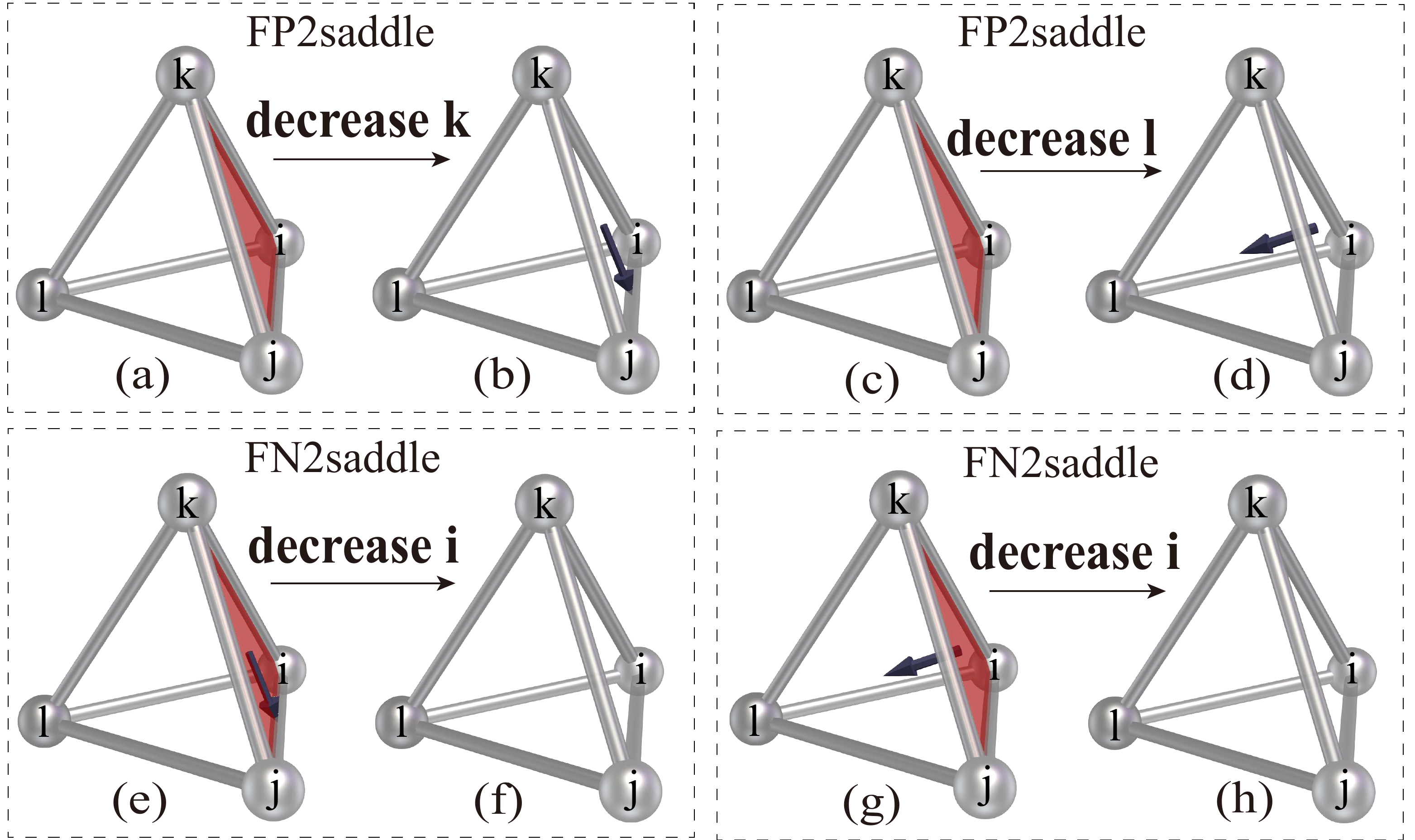}
    \caption{Fixing an FP2saddle that should be paired with an edge, as shown in (a) and (b), or with a tetrahedron, as shown in (c) and (d); and an FN2saddle that is incorrectly paired with an edge, as shown in (e) and (f), or with a tetrahedron, as shown in (g) and (h).}
    \label{fig:2saddles} 
    \vspace{-0.8em}
\end{figure}

\noindent \textbf{False Positive 2-Saddle (FP2saddle)}. An FP2saddle triangle $ijk$ in the original data should be paired with an edge/tetrahedron in one of the two conditions: 1. paired with edge $ij$, requiring $f_i > f_k$ and $f_j > f_k$; 2. paired with tetrahedron $ijkl$, requiring $ f_i > f_l $ and $ f_j > f_l $ and $ f_k > f_l $. We can edit one of the vertices associated with the edge or tetrahedron paired with triangle $ijk$ in the original data, as shown in Figure~\ref{fig:2saddles}. 

\noindent \textbf{False Negative 2-Saddle (FN2saddle).} Similar to the FN1saddle cases, if an FN2saddle is incorrectly paired with an edge $ij$ then we edit vertex $i$ or $j$ (as shown in Figure~\ref{fig:2saddles}(e) and (f)); if it is paired with a tetrahedron $ijkl$, then we edit vertex $i$ or $j$ or $k$ (as shown in Figure~\ref{fig:2saddles}(g) and (h)).

\begin{figure}[htb!]
    \centering
    \includegraphics[width=\linewidth]{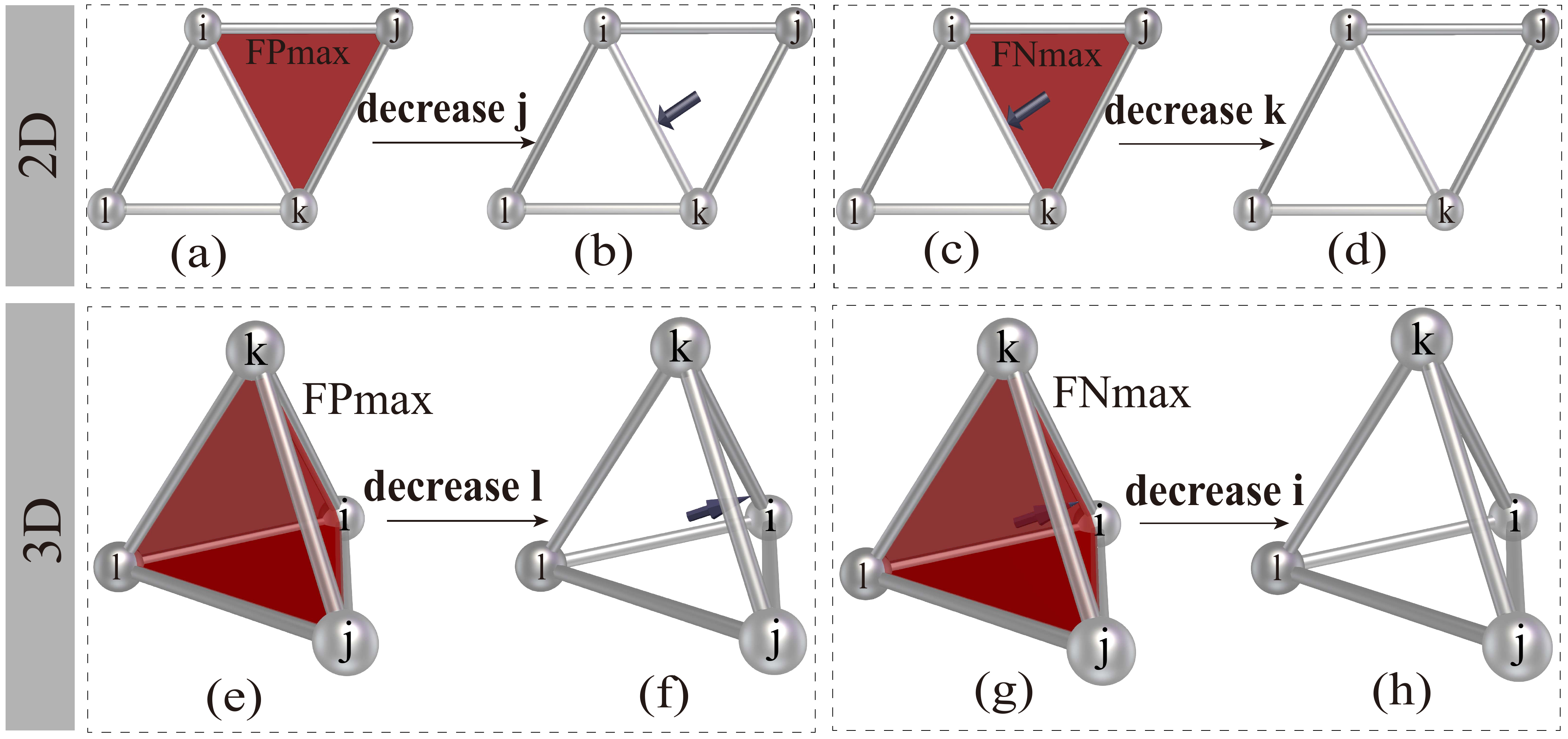}
    \caption{Fixing an FPmax in 2D, as shown in (a) and (b), and in 3D, as shown in (e) and (f); and an FNmax in 2D, as shown in (c) and (d), and in 3D, as shown in (g) and (h).}
    \label{fig:max}
    
\end{figure}

\noindent\textbf{False Positive Maximum (FPmax)}. In 2D, a maxima is represented by a triangle, while in 3D, it is represented by a tetrahedron. Therefore, we discuss these cases separately. In a 2D case, as shown in Figure~\ref{fig:max}(a) and (b), triangle $ijk$ is an FPmax which is paired with edge $ik$ in the original data (as shown in Figure~\ref{fig:max}(b)), meaning the scalar values of vertex $i$, vertex $k$, and vertex $j$ must satisfy that $f_i > f_j$ and $f_k > f_j$. We then decrease the value of vertex $j$ until this condition is also satisfied in the decompressed data. In a 3D case, the FPmax is a tetrahedron $ijkl$, as shown in Figure~\ref{fig:max}(e) and (f). Assuming that tetrahedron $ijkl$ is paired with a triangle $ijk$ in the original data. Similarly, we decrease vertex $l$.

\noindent\textbf{False Negative Maximum (FNmax)} could also be divided into two cases:  
In a 2D case, the FNmax is a triangle $ijk$, as shown in Figure~\ref{fig:max}(c). Assuming that triangle $ijk$ is paired with an edge $ik$ in the decompressed data, this implies that $\hat{f}_i > \hat{f}_j$ and $\hat{f}_k > \hat{f}_j$. We decrease vertex $i$ or $k$ to break this condition, leaving triangle $ijk$ unpaired with any edge. In a 3D case, the FNmax is a tetrahedron $ijkl$ as shown in Figure~\ref{fig:max}(g). Assuming that tetrahedron $ijkl$ is paired with a triangle $ijk$ in the decompressed data, we decrease vertex $i$, $j$, or $k$ to leave tetrahedron $ijkl$ unpaired.

\subsection{Separatrices Loops (S-Loops)}

After all false critical cells are fixed, we proceed to fix the incorrect gradient pairings encountered when tracing separatrices from each saddle. Each iteration of our method consists of three steps: (1) in the $k$th iteration, extract separatrices from the currently edited data $g^{(k)}$ based on gradient pairings;
(2) identify the corresponding \emph{troublemaker};
(3) edit the scalar value of the vertices forming the \emph{troublemaker} to correct the pairing.

The troublemaker may be a 0-cell (paired with an incorrect 1-cell in descending separatrices), a 1-cell (paired with an incorrect 2-cell in saddle-saddle connectors), or a 2-cell (paired with an incorrect 3-cell in ascending separatrices). We now analyze each case based on the different dimensions of the troublemaker.

\begin{figure}[htb!]
    \centering
    \includegraphics[width=\linewidth]{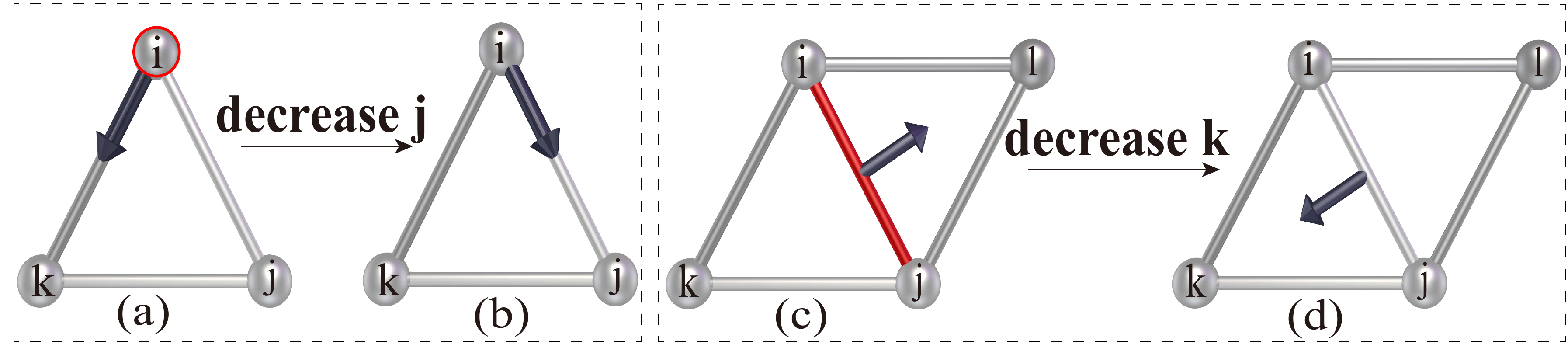}
    \caption{Illustration of fixing troublemakers in 2D: (a) and (b) show cases where the troublemaker is a vertex, while (c) and (d) show cases where it is an edge.}
    \label{fig:1dtm} 
    
\end{figure}

A \textbf{0-dimensional} troublemaker vertex $i$, as shown in Figure~\ref{fig:1dtm}(a) and (b), is incorrectly paired with edge $ik$ in the decompressed data, whereas vertex $i$ should be paired with edge $ij$ in the original data. According to the gradient pairing rule, vertex $i$ should pair with the edge in $P_{i}$ that has the smallest function value. To correct this, we decrease vertex $j$, ensuring that vertex $i$ pairs correctly with edge $ij$.

A \textbf{1-dimensional} troublemaker edge $ij$ should pair with triangle $ijk$ in the original data but is incorrectly paired with triangle $ijl$ in the decompressed data (as shown in Figure~\ref{fig:1dtm}(c) and (d), meaning that the function value of triangle $ijk$ is greater than that of triangle $ijl$ in $\hat{f}$. Note that triangles $ijl$ and $ijk$ share vertices $i$ and $j$, so their scalar ordering is determined by the relative values of vertices $k$ and $l$. We decrease vertex $k$, making edge $ij$ pair correctly with triangle $ijk$.

The process is similar for a \textbf{2-dimensional} troublemaker triangle $ijk$. Suppose tetrahedron $ijkl$ is the tetrahedron paired with triangle $ijk$ in the original data, and in the decompressed data, triangle $ijk$ is paired with tetrahedron $ijkq$. For a similar reason as in the 2D troublemaker case, we decrease vertex $l$.

\subsection{Convergence Analysis}\label{sec:convergence}
Our workflow alternates between C-loops and S-loops to iteratively fix false critical cells and separatrices. Note that edits made during the C-loops may introduce new distortions in separatrices, while edits in the S-loops may create new false critical cells, we must iteratively execute C-loops and S-loops to address these newly introduced false cases. To ensure that this alternating procedure eventually terminates with the fully preserved MSC, we analyze the convergence of our method by first proving that each individual loop terminates and then demonstrating the convergence of the overall workflow.

\textbf{Convergence of C- and S-Loops} The false critical cells in the C-loops are caused by incorrect gradient pairing results, which are determined by the scalar value order of the vertices constituting each cell. Modifying the scalar values of the vertices modifies the scalar values of the associated cells, thereby correcting the gradient pairing results. However, modifying the scalar value of any vertex affects all cells containing the vertex, potentially introducing new false cases and requiring further iterations of the C-loop.

The iterative process of the C-loop is guaranteed to converge because gradient pairings depend solely on the scalar value order of the vertices. Since all edits strictly decrease vertex values, and each vertex has a lower bound at $f_i - \xi$, no vertex can be edited infinitely many times. Even if a modification introduces new false critical cells, the total number of modifications is finite. In the worst case, all vertices reach their lower bound, at which point their relative order matches that of the original data. Once the scalar order is recovered, gradient pairing results are guaranteed to match those of the original data, eliminating all false critical cells. Therefore, the C-loop converges after a finite number of iterations. Similarly, modifications in the S-loops also strictly decrease vertex scalar values toward their lower bounds, ensuring the convergence of the S-loops for the same reason.

\textbf{Convergence of Alternating C- and S-Loops.}
S-loops may introduce new false critical cells by modifying vertex values along separatrices. However, these new false critical cells are always corrected in subsequent C-loops. Since all edits strictly decrease vertex values and remain bounded, this alternating process cannot continue indefinitely.
At worst, all vertices reach their lower bounds, ensuring that their relative scalar order is fully determined, restoring gradient pairings to match the original data. As a result, all false critical cells and separatrix distortions are eventually resolved, guaranteeing convergence within a finite number of iterations.

\subsection{Multitier Preservation}

We designed our method to support multiple tiers of topological feature preservation to accommodate different application needs while balancing efficiency and topology preservation. Our method allows early termination based on the selected tier. For example, T1 and T2 focus on preserving extrema or critical cells. In these cases, only C-loops are needed, and the process stops once all false extrema or critical cells are fixed. T3 also preserves the connectivity between saddles and extrema, so S-loops are used to detect and fix incorrectly connected critical cells. T4 further requires accurate gradient paths, which means more troublemakers must be identified and corrected compared to T3.

\subsection{Quantized Representation of Edits}\label{sec:edits}

To reduce the storage overhead associated with storing edits, we propose a quantized representation that efficiently encodes edit values while maintaining feature preservation, while MSz directly stores the edits $\delta_i$ as floating-point numbers using a key-value pair format. However, storing edits in floating-point representation incurs significant storage costs, particularly for large datasets or frequent edits.

In contrast, our method applies a fixed edit step of $\xi/2^{q_{\max}}$ at each iteration, as defined in Equation~\eqref{eq1}. The quantized form allows us to store only the integer count of edits: $q$ for each vertex instead of floating-point values. The total edit at a vertex is simply computed as $\delta_i = q \cdot \xi/2^{q_{\max}}$, reducing storage overhead while ensuring the exact reconstruction of edited values.

Furthermore, we apply an additional compression to the quantized edits. Since $q_{\max}$ is small (e.g., from 2 to 6, as shown in our experiments), each quantized edit can be represented using a fixed number of bits. We can convert the list of edits into a compact bitstream, which we further compress using a dictionary-based algorithm such as ZSTD~\cite{zstd}.

\subsection{GPU Implementation with Reduced Memory Footprint}

We implement our method using data-parallel GPU kernels to enable efficient processing of large scientific datasets, where gradient pairing and false-case handling are performed independently across cells.
A key challenge is the high memory footprint of a straightforward GPU implementation, which can exceed the capacity of modern GPUs for large grids.
To address this challenge, we leverage two key techniques, \emph{gradient pairing via lookup tables} and \emph{kernel fusion}, as discussed below.

\subsubsection{Gradient Pairing via Lookup Tables}
Our method requires storing two copies of gradient-pairing results for all simplices during GPU execution for both original data and decompressed data.
For a regular 3D grid subdivided into six tetrahedra per cube, the simplicial complex contains approximately
$N_0$ vertices, $N_1\!\approx\!7N_0$ edges, $N_2\!\approx\!12N_0$ triangles, and $N_3\!\approx\!6N_0$ tetrahedra,
i.e., about $2 \times (N_0+N_1+N_2+N_3) \approx 52N_0$ simplices in total.

To reduce this overhead, we encode gradient pairing results using \textbf{dimension-specific lookup tables} that exploit the regular structure of the grid, without explicitly storing global cell indices.
Each lookup table compactly represents local incidence relations (e.g., vertex-edge, edge-triangle, or triangle-tetrahedron pairings) using 8-bit codes.
During execution, these codes are decoded on the fly to recover the corresponding global simplex indices, eliminating the need to store explicit index lists.

The reduction in memory usage introduces a moderate increase in end-to-end runtime caused by the need to decode each lookup-table entry back into the corresponding global simplex indices. 
A detailed quantitative analysis of the resulting runtime--memory trade-off, together with the formal definition of the lookup tables, is provided in Appendix~E.

\subsubsection{Kernel Fusion for False-Case Handling}
In a naive GPU implementation, false-case extraction and correction are performed in separate kernels, requiring intermediate buffers to store all detected false cases.
We reduce memory usage by fusing false-case detection and correction into a single kernel, allowing false cases to be processed immediately without materializing intermediate arrays.

\subsubsection{Resolving Read-Write Conflicts}
False-case correction is performed in parallel across cells, which may lead to concurrent updates to shared simplices (e.g., vertices).
We resolve the read-write conflicts using atomic operations, ensuring that at most one correction modifies a given vertex at a time while preserving parallel execution.

\section{Evaluation}

We evaluate our method on an NVIDIA A100 GPU using datasets from various fields (as summarized in Table~\ref{tab:memory_tradeoff}). SZ3, ZFP, SPERR, and MGARD are selected as the base compressors to generate the decompressed data. All datasets are defined on simply connected domains with regular grid-based triangulation, which aligns with our current implementation. We evaluate different values of $q_{\max}$ and select $q_{\max} = 6$ based on a trade-off between storage and computational overhead. Unless otherwise noted, all reported results are measured per timestep.

\begin{table}[htb!]
\centering
\small
\renewcommand{\arraystretch}{1.25}
\setlength{\tabcolsep}{1.8pt}
\caption{
Datasets used in our evaluation, along with spatial dimensions, timesteps, and GPU memory usage.
}
\label{tab:memory_tradeoff}
\begin{tabular}{lcccc}
\toprule
\textbf{Dataset} 
& \textbf{Dimensions} 
& \textbf{Timesteps}
& \begin{tabular}{c}\textbf{GPU Memory}\\\textbf{Usage (GB)}\end{tabular} \\
\midrule
AT           
& $177\times95\times48$     
& 1     
& \textbf{0.57}  \\

Heated Flow  
& $150\times450\times1$            
& 2001  
& \textbf{0.46}  \\

IVT  
& $576\times361\times1$            
& 3830
& \textbf{0.47}  \\

Vortex       
& $128\times128\times128$   
& 98  
& \textbf{0.72}  \\

CESM         
& $3600\times1800\times1$ 
& 79  
& \textbf{0.70}  \\

Turb. Comb.
& $480\times720\times120$ 
& 122  
& \textbf{5.68}  \\

Deep Water 
& $460\times280\times240$ 
& 50  
& \textbf{4.36}  \\

Miranda      
& $256\times384\times384$   
& 7  
& \textbf{5.23}  \\

S3D          
& $500\times500\times500$   
& 11  
& \textbf{16.24} \\

NYX          
& $512\times512\times512$   
& 6  
& \textbf{17.6}  \\
\bottomrule
\end{tabular}
\end{table}

We highlight the experimental results in four aspects: 1. DMTz preserves the full MSC under all tested error bounds, whereas the base compressors (e.g., SZ3 and ZFP) require extremely tight error bounds to preserve the MSC, and the magnitude of topological distortion introduced by lossy compression increases with the error bound, as reflected by the number of required edits; 2. For preservation tiers from T1 through 4, the computation and storage overhead generally grow with the level of preservation. 3. Our quantized edit strategy reduces the storage overhead compared to MSz~\cite{li_msz} by up to 3.4$\times$ in our experiments.
4. Our GPU-accelerated implementation achieves up to 26 $\times$ speedup for gradient pairing compared with using all 64 threads on the CPU node.

\subsection{Evaluation Metrics}
We evaluate our method using several key metrics. The \textbf{Critical/Separatrices Recalls} quantify the proportion of critical cells and separatrices in the original dataset that are successfully retained in the decompressed data. \textbf{Critical/Separatrices Precisions} quantify the proportion of detected critical cells and separatrices in the decompressed data that also exist in the original data. The $L^2$-\textbf{Wasserstein Distance ($W_2$)}~\cite{Edelsbrunner_10} represents the topological difference between the persistence diagram derived from the MSCs of the original and decompressed data, with lower values indicating better preservation of persistent features. The \textbf{Compression Ratio (CR)} measures the efficiency of compressing the original scalar field alone, computed as the size of the original data divided by the size of the compressed data. The \textbf{Overall Compression Ratio (OCR)} quantifies the compression efficiency after incorporating compressed edits. It is computed as the original data size divided by the combined size of compressed edits and data. The \textbf{Edit Ratio} measures the proportion of modified data points necessary to fully preserve the MSC in the decompressed data and is calculated as the number of modified points divided by the number of data points. 

\begin{figure*}
    \centering
    \includegraphics[width=\linewidth]{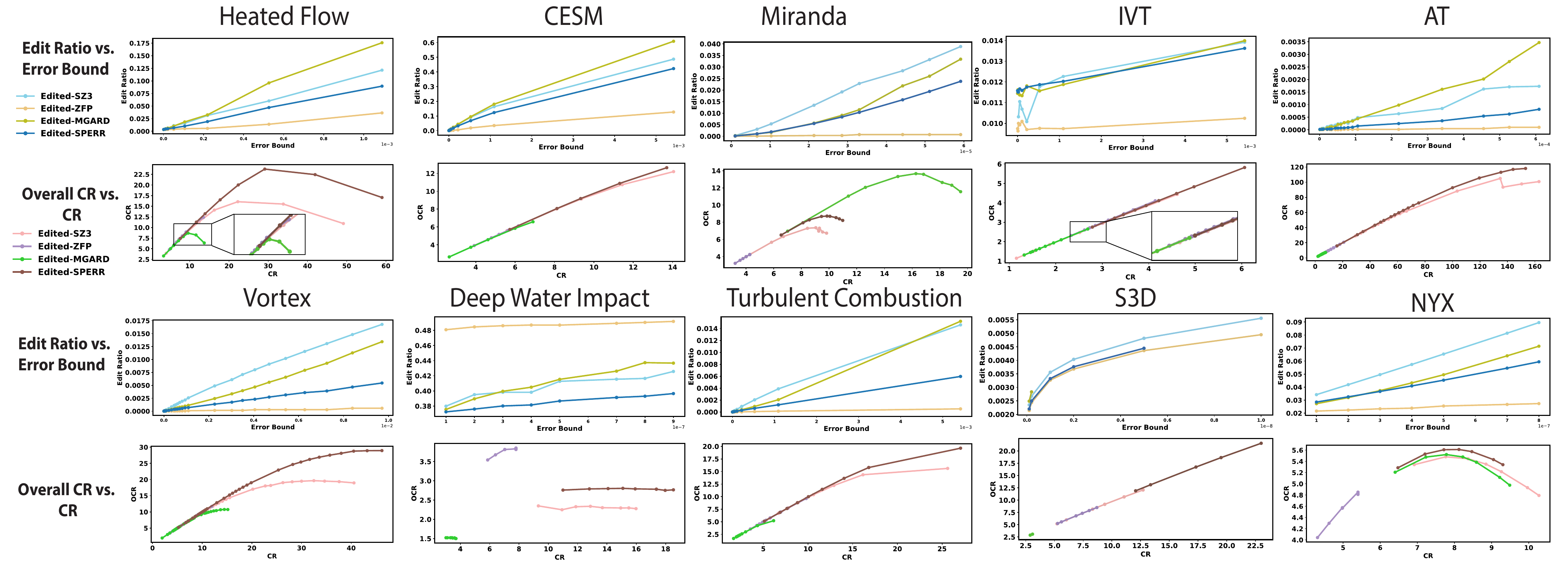}
    \caption{\updated{Comparison of different lossy compressors (SZ3, ZFP, MGARD, and SPERR) in their ability to preserve Morse--Smale complexes (MSCs) across datasets (per timestep) under varying error bounds.}}
    \label{fig:bit_error}
    \vspace{-1em}
\end{figure*}

\subsection{Comparison of Different Base Compressors with Different Error Bounds}

\begin{mdframed}
\noindent\textbf{Summary.}
Across all datasets and error bounds, DMTz 100\% preserves MSC, whereas base compressors alone fail to preserve the correct topology.
\end{mdframed}

We demonstrate the ability of our method to fully preserve the MSC with different compressors and error bounds across various datasets while maintaining acceptable storage overhead, as shown in Figure~\ref{fig:bit_error}. 
The number of edits generally shows a proportional relationship with the error bounds. We also observed that the number of required edits increases with the complexity of the dataset; for instance, the average edit ratio is lower for the Adenine Thymine (AT) dataset compared to the other datasets.

\updated{Comparing different base compressors (SZ3, ZFP, MGARD, and SPERR), we observe distinct topological distortion patterns tied to their underlying algorithms. These patterns directly dictate the required edits, correction overhead size, and convergence time. For instance, SZ3 relies on local prediction; while it often achieves the highest compression ratio (CR), it can introduce localized prediction errors that invert local gradients, resulting in more required topological edits and longer convergence times. In contrast, ZFP uses an orthogonal block transform and generally yields lower CRs at equivalent error bounds, retaining higher baseline data fidelity, thus introducing fewer initial topological distortions, requiring fewer iterations and smaller edit overhead. MGARD, built on multigrid hierarchical interpolation, exhibits a noticeably higher edit ratio, particularly on structurally simpler datasets such as AT and Heated Flow. SPERR, using discrete wavelet transforms, generally demonstrates moderate behavior, yielding intermediate values for both required edit ratio and overall CR.
An exception is observed in the Deep Water Impact dataset, where ZFP achieves a higher overall CR than other compressors. This occurs because the dataset features extensive flat regions where SZ3 flattens subtle gradients, triggering massive topological distortions and required edits. Meanwhile, MGARD's low base CR combined with high edit volume results in the lowest overall CR.}

\updated{To further quantify the impact of different base compressors, Table~\ref{tab:compressor_comparison} compares the number of iterations and edit sizes under the same error bound ($10^{-4}$). The results show that compressors inducing stronger local distortions generally require more iterations and larger edit overhead. For example, SZ3 tends to introduce local prediction errors that disrupt gradient consistency, leading to higher iteration counts and edit sizes, while ZFP preserves more global structure and therefore requires fewer corrections. MGARD often incurs larger overhead due to its hierarchical approximation introducing widespread deviations, whereas SPERR exhibits intermediate behavior. }

\updated{Overall, these results further show that the correction cost of DMTz is influenced by the type of compression-induced distortions. Compressors that better preserve scalar ordering reduce both the number of iterations and the edit overhead, while those introducing localized or hierarchical distortions lead to more extensive and iterative corrections.}

\begin{table}[htb!]
\centering
\caption{\updated{Comparison of DMTz performance (iteration numbers and edit size) across base compressors at $10^{-4}$ error bound. Iter. denotes the number of iterations; Edit Size represents the compressed edit overhead in MB.}}
\label{tab:compressor_comparison}
\small
\renewcommand{\arraystretch}{0.9}
\setlength{\tabcolsep}{2pt}
\begin{tabular}{l cc cc cc cc}
\toprule
\multirow{2}{*}{\textbf{Dataset}} & \multicolumn{2}{c}{\textbf{SZ3}} & \multicolumn{2}{c}{\textbf{ZFP}} & \multicolumn{2}{c}{\textbf{MGARD}} & \multicolumn{2}{c}{\textbf{SPERR}} \\
\cmidrule(lr){2-3} \cmidrule(lr){4-5} \cmidrule(lr){6-7} \cmidrule(lr){8-9}
& Iter. & \begin{tabular}{c}Edit\\Size\end{tabular} & Iter. & \begin{tabular}{c}Edit\\Size\end{tabular} & Iter. & \begin{tabular}{c}Edit\\Size\end{tabular} & Iter. & \begin{tabular}{c}Edit\\Size\end{tabular} \\
\midrule
Heated Flow          & 111 & 0.003 & 60 & 0.0003 & 110 & 0.002 & 83 & 0.002 \\
IVT          & 74 & 0.006 & 56 & 0.001 & 67 & 0.004 & 68 & 0.006 \\
AT                   & 53 & 0.002 & 9 & 0.0001 & 76 & 0.003 & 41 & 0.001 \\
CESM                 & 1207 & 0.321  & 247 & 0.028 & 625 & 0.17 & 811 & 0.15 \\
Vortex               & 19 & 0.004 & 9 & 0.0007 & 19 & 0.002 & 14 & 0.001 \\
Deep Water    & 669 & 51.07 & 569 & 45.81 & 1237 & 100.3 & 1033 & 97.82 \\
Miranda              & 975 & 21.31 & 804 & 0.67 & 1696 & 18.50 & 1329 & 13.93 \\
Turb. Comb. & 47 & 0.121  & 15 & 0.002 & 38 & 0.062 & 41 & 0.046\\
S3D                  & 611 & 1.022  & 386 & 0.53 & 875 & 1.072 & 506 & 0.938 \\
NYX                  & 155 & 26.29 & 116 & 9.75 & 133 & 20.88 & 139 & 21.30 \\
\bottomrule
\end{tabular}
\end{table}

\subsection{Computation and Storage Overhead}\label{section:edits} 
\begin{mdframed}
\noindent\textbf{Summary.}
Increasing the preservation tier increases both computation and storage costs; however, compared with the floating-point strategy used in prior work, our quantized edits achieve an approximately 93.9\% reduction in storage (3366 bytes to 205 bytes) in our experiment.
\end{mdframed}
\begin{figure}[htb!]
    \centering
    \includegraphics[width=\linewidth]{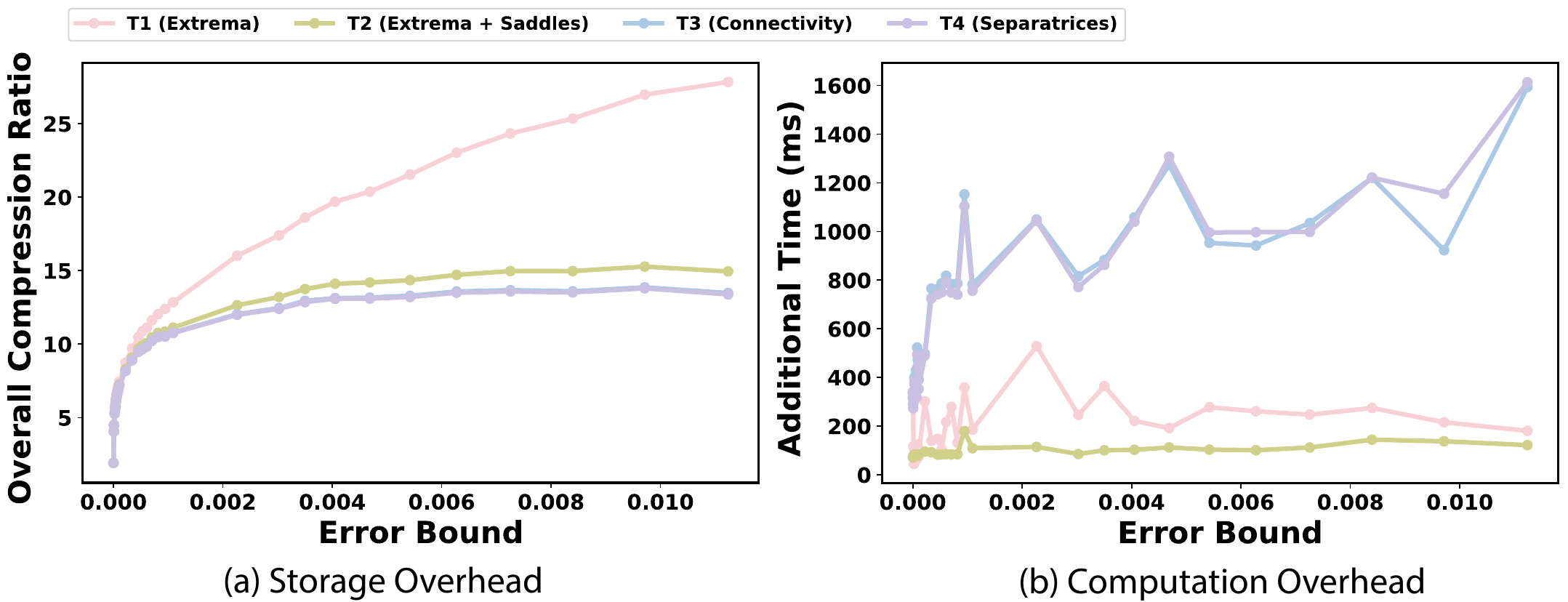}
    \caption{Storage and computational overhead of different preservation tiers on the IVT dataset using SZ3: T1 (pink), T2 (brown), T3 (blue), and T4 (purple). The T3 and T4 curves overlap in (a).}
    \label{fig:computation_overhead}
\end{figure}

\begin{figure}[htb!]
    \centering
    \includegraphics[width=\linewidth]{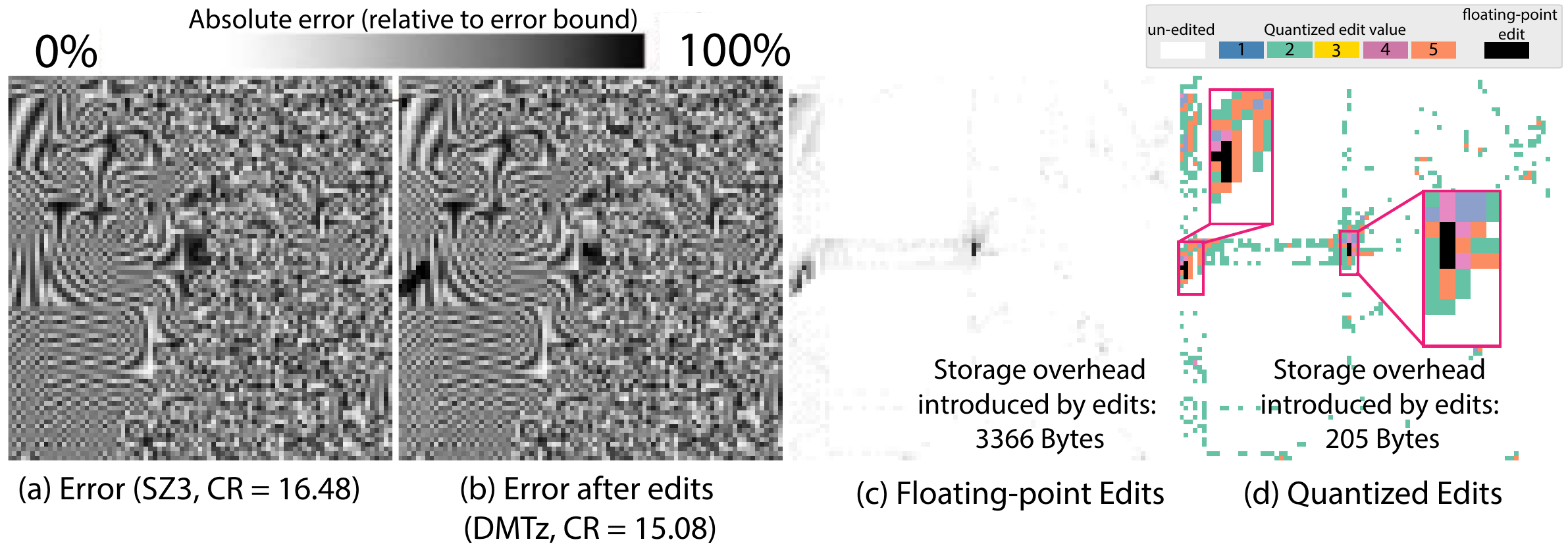}
    \caption{Error maps and edit distributions for a cropped region of the Heated Flow dataset. (a) Error map of SZ3’s decompressed data (relative error bound $= 10^{-4}$). (b) Error after applying edits (DMTz). (c) Edit values applied to SZ3’s decompressed data. (d) Distribution of quantized edits.}
    \label{fig:error} 
    
\end{figure}
To evaluate the overhead introduced by our method as the preservation tier increases, we analyze the computation and storage costs across different tiers using the Integrated Vapor Transport (IVT) dataset. The IVT dataset is an ensemble of reanalysis fields that represent the amount of water vapor transported horizontally across each grid point per unit time. 
We chose the IVT dataset because of its complex topological structure, which presents meaningful challenges for preserving multitier topological features.

As the preservation tier increases, both computation and storage overheads generally increase, as shown in Figure~\ref{fig:computation_overhead}. This trend is expected, as finer preservation imposes stricter constraints on allowable data changes, thus requiring more edits and iterations.

Furthermore, the quantized edits reduce the storage overhead by quantizing the majority of edits, compared to the floating-point edits used in MSz~\cite{li_msz}. As illustrated in Figure~\ref{fig:error}(c) and (d), the quantization of edits reduces the storage overhead from 3366 bytes to 205 bytes in the Heated Flow dataset.

\subsection{Comparison of our method and MSz~\cite{li_msz}}~\label{sec:msz_cp}
\begin{mdframed}
\noindent\textbf{Summary.}
In our experiment, MSz incurs a total of 588 saddle errors and 508 maximum errors (comprising both false positives and false negatives), whereas DMTz achieves 100\% MSC preservation, using about 3.4$\times$ less storage (66.62\,KB to 19.57\,KB) and only a 2$\times$ computation overhead (0.01\,s to 0.02\,s).
\end{mdframed}

Our method differs fundamentally from MSz in its preservation goal. 
MSz only preserves PLMSS, which serves as a computationally cheaper preview of the full MSC, but does not preserve saddle points or saddle-related structures, lacking the structural completeness needed for further topological analysis, such as simplification, as MSC can do. This limitation results in topological inconsistencies such as false saddles and incorrect connections between critical points. Therefore, preserving PLMSS is a fundamentally different goal and is insufficient for ensuring the correctness of the MSC. In contrast, DMTz preserves the full MSC, guaranteeing the preservation of all critical points (including saddles) and ensuring the correctness of separatrices, directly addressing the key limitations of MSz.
Preserving the full MSC is significantly more challenging than preserving PLMSS. MSz detects and fixes false cases through local vertex comparisons, making the process relatively simple. DMTz, by contrast, computes discrete gradient pairings across cells of different dimensions (e.g., vertex-edge or edge-triangle), introducing complex dependencies that make preservation more difficult.

\begin{figure}[htb!]
    \centering
    \includegraphics[width=\columnwidth]{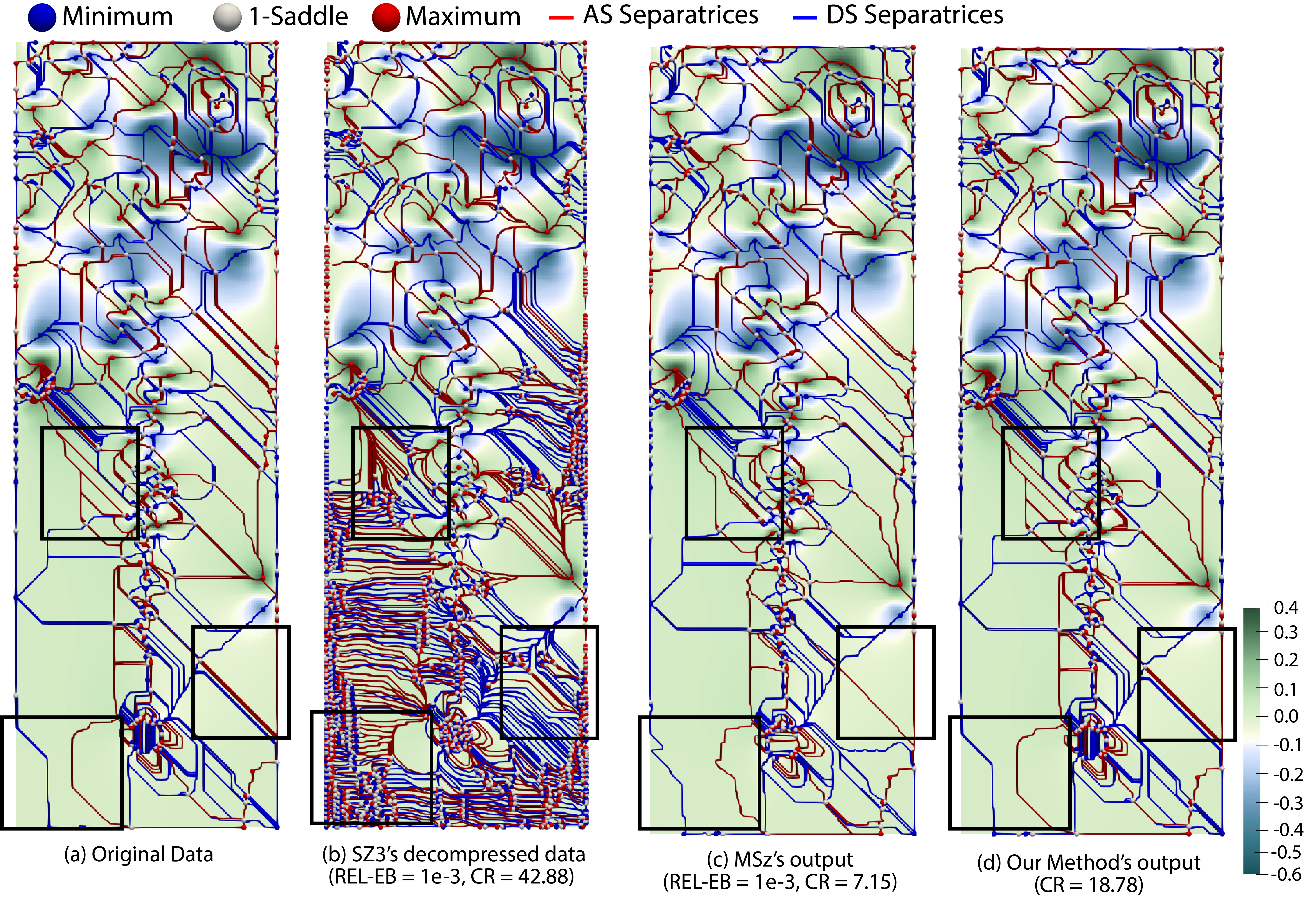}
    \caption{Critical points and separatrices of the Heated Flow dataset: (a) original data, (b) SZ3’s decompressed data (relative error bound $= 10^{-3}$), (c) MSz’s output, and (d) our method’s output. Black boxes highlight regions with structural differences.}
    \label{fig:Msz_comparison}
    \vspace{-1em}
\end{figure}

To evaluate the difference between MSz and DMTz (ours), we compare the output of DMTz with that of MSz, using SZ3 as the base compressor with a relative error bound of $10^{-3}$ on the heated flow dataset (150 × 450 double-precision values, 0.51 MB), as shown in Figure~\ref{fig:Msz_comparison}. While MSz preserves PL local extrema, it still introduces distortions in saddles and separatrices in the discrete scalar field (as shown in the black boxes). This comparison evaluates two key aspects: topological preservation, storage and computational overhead (see Table~\ref{tab:msz_vs_dmtz}). 

\textbf{Topological Preservation.} 
Although MSz produces a similar number of critical points compared to the original data (98 minima, 934 saddles, and 837 maxima versus 98 / 703 / 606), a large portion of these critical points are incorrect. In particular, MSz introduces 588 false saddles, indicating significant distortion in the topological structure. Moreover, the separatrices are heavily degraded (Figure~\ref{fig:Msz_comparison}(c)), recovering only 47\% of the original separatrices, whereas our method preserves 100\%.

\textbf{Storage and computational overhead.} MSz introduces a storage overhead of 66.62KB, which is 3.4 times larger than that of our method (19.57KB), because MSz stores edits in floating-point format. On the other hand, MSz incurs a slightly lower computational overhead (0.01s) compared to our method (0.02s). However, our method fully preserves all critical points and separatrices, which involve significantly more complex topological structures than those considered in MSz. Despite the acceptable increase in cost, our method remains meaningful for applications requiring topological accuracy.

\begin{table}[htb!]
    \centering
    \caption{Comparison of MSz and DMTz (ours). The inputs are compressed under a relative error bound of $10^{-3}$}
    {\small
    \renewcommand{\arraystretch}{1} %
    \setlength{\tabcolsep}{1pt} %
    \begin{tabular}{lccc}
        \toprule
        & \textbf{Original} & \textbf{MSz} & \textbf{DMTz (Ours)}  \\
        \midrule

        Min / Saddle / Max  & 98 / 703 /606 & 98 / 934 / 837 & 98 / 703 / 606 \\
        False critical cells & - & 0 / 
 588 / 508 & \textbf{0 / 0 / 0}  \\
        Separatrices Recall & 1.00 & 0.47 & \textbf{1.00}  \\
        Separatrices Precision & 1.00 & 0.47 & \textbf{1.00}  \\
        Critical Recall & 1.00 & 0.66 & \textbf{1.00}  \\
        Critical Precision & 1.00 & 0.64 & \textbf{1.00}  \\
        Runtime (s) & - & 0.01 & 0.02  \\
        Compression Ratio & - & 7.15 & 16.26 \\
        \bottomrule
    \end{tabular} 
    }
    \label{tab:msz_vs_dmtz}
\end{table}

\subsection{Compression of Edits}
\begin{mdframed}
\noindent\textbf{Summary.}
Quantized edits deliver up to 3.2$\times$ higher overall compression ratios compared with MSz, and $q_{\max}$ introduces a balance between compression ratio and computational overhead.
 \end{mdframed}

\begin{figure}[htb!]
    \centering
    \includegraphics[width=\linewidth]{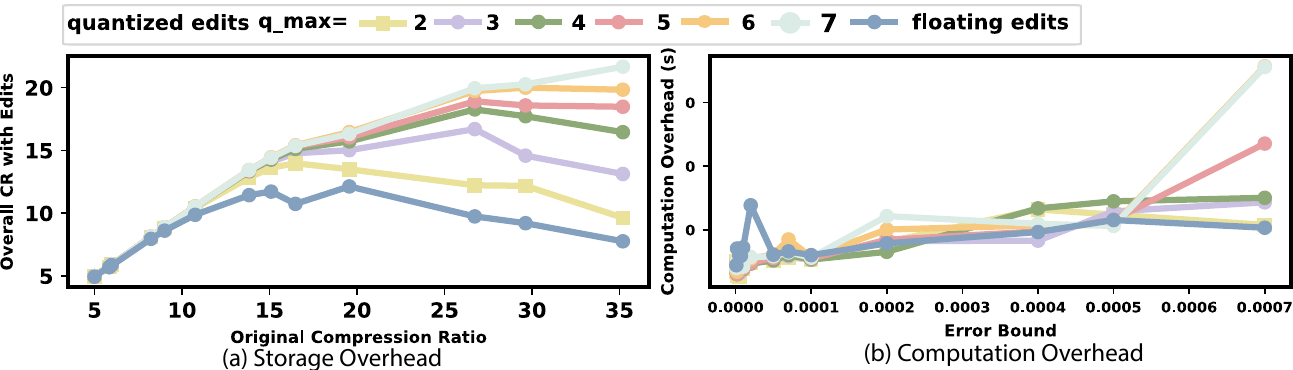}
    \caption{(a) Storage overhead and (b) computational overhead induced by edits on the Heated Flow dataset using SZ3 as the compressor, with $q_{\max}$ ranging from 2 to 7.} 
    \label{fig:edit_comparison}
    
\end{figure}

We test a range of relative error bounds from $10^{-6}$ to $10^{-2}$ on the heated flow dataset, using SZ3 as our base compressor (as shown in Figure~\ref{fig:edit_comparison}). Our method achieves up to 3.2$\times$ higher compression ratios than MSz~\cite{li_msz}, with improvements becoming more pronounced at higher error bounds, because larger error bounds tend to introduce greater topological distortions that require more edits. By quantizing most of the edits, our method reduces the storage overhead. We also observe that increasing $q_{\max}$ improves the compression ratio by allowing more edits to be quantized. However, while compression improves substantially when increasing $q_{\max}$ from 2 to 6, the improvement slows down beyond $q_{\max} = 6$, even as computational overhead still increases. Based on this trade-off, we selected $q_{\max} = 6$ in our experiments. Note that the choice of $q_{\max}$ is data-dependent; a formal study of optimal $q_{\max}$ selection is left for future work. A more complete evaluation of different $q_{\max}$ across all datasets is provided in Appendix D.

\begin{table*}[htb!]
\centering
\caption{\updated{Comparison between Floating Edits and Quantized Edits across different datasets. SZ3 is used as the base compressor with an error bound of $10^{-4}$. Note that Raw Size, Compressed Size, and CR specifically report the overhead of the topological edits, while OCR denotes the Overall Compression Ratio. Edits are losslessly compressed by ZSTD.}}
\label{tab:edits_comparison}
\small
\renewcommand{\arraystretch}{0.8}
\setlength{\tabcolsep}{0.7pt}
\begin{tabular}{l cccccc cccccc}
\toprule
\multirow{2}{*}{\textbf{Dataset}} & \multicolumn{6}{c}{\textbf{Floating Edits}} & \multicolumn{6}{c}{\textbf{Quantized Edits}} \\
\cmidrule(lr){2-7} \cmidrule(lr){8-13}
& \begin{tabular}{c}Raw Size\\(MB)\end{tabular} 
& \begin{tabular}{c}Compressed\\Size (MB)\end{tabular} 
& \begin{tabular}{c}CR \\ (Edits)\end{tabular} 
& \begin{tabular}{c}OCR\end{tabular}
& \begin{tabular}{c}Correction\\Time (s)\end{tabular} 
& \begin{tabular}{c}Correction\\Thrpt. (MB/s)\end{tabular} 
& \begin{tabular}{c}Raw Size\\(MB)\end{tabular} 
& \begin{tabular}{c}Compressed\\Size (MB)\end{tabular} 
& \begin{tabular}{c}CR \\ (Edits)\end{tabular} 
& \begin{tabular}{c}OCR\end{tabular}
& \begin{tabular}{c}Correction\\Time (s)\end{tabular} 
& \begin{tabular}{c}Correction\\Thrpt. (MB/s)\end{tabular} \\
\midrule
Heated Flow          & 0.015 & 0.010 & 1.45 & 11.94 & 1.08 & 0.48   & 0.0037 & 0.0035 & 1.05 & \textbf{14.30} & 0.09 & \textbf{5.72} \\
IVT                  & 0.024 & 0.014 & 1.66 & 7.49  & 0.07 & 22.66  & 0.008 & 0.006 & 1.33 & \textbf{8.50}  & 0.05 & \textbf{31.73} \\
AT                   & 0.012 & 0.008 & 1.39 & 59.21 & 0.16 & \textbf{38.48}  & 0.0027 & 0.0024 & 1.08 & \textbf{63.07} & 0.37 & 16.64 \\
CESM                 & 1.52  & 1.01  & 1.50 & 10.75 & 6.27 & 7.88   & 0.37 & 0.32 & 1.15 & \textbf{12.64} & 3.83 & \textbf{12.90} \\
Vortex               & 0.015 & 0.011 & 1.44 & 7.31  & 1.44 & \textbf{11.11}  & 0.0049 & 0.0041 & 1.18 & \textbf{7.93}  & 1.86 & 8.60 \\
Deep Water    & 114.2 & 63.28 & 1.89 & 2.37  & 422.58 & 0.558  & 96.65 & 51.07 & 1.89 & \textbf{2.85} & 290.1 & \textbf{0.812} \\
Miranda              & 46.62 & 30.33 & 1.53 & 5.23  & 249.1 & 1.15   & 23.38 & 21.31 & 1.06 & \textbf{6.25} & 215.5 & \textbf{1.33} \\
Turb. Comb. & 0.62  & 0.41  & 1.50 & 8.67  & 136.8 & \textbf{2.31}   & 0.16 & 0.12 & 1.31 & \textbf{8.94} & 138.8 & 2.27 \\
S3D                  & 3.99  & 2.29  & 1.74 & 4.91  & 814.12 & \textbf{1.17}   & 2.37 & 1.02 & 2.30 & \textbf{5.20} & 824.57 & 1.15 \\
NYX                  & 52.98 & 36.64 & 1.44 & 3.00  & 710.89 & 1.44   & 28.75 & 26.29 & 1.09 & \textbf{3.15} & 189.28 & \textbf{5.41} \\
\bottomrule
\end{tabular}
\end{table*}

\subsection{Analysis of Quantized vs. Floating Edits}
\label{app:quantized_analysis}
\updated{Compared to floating edits used in MSz~\cite{li_msz}, quantized edits consistently reduce the storage overhead by representing correction values more compactly (Table~\ref{tab:edits_comparison}), leading to smaller edit sizes (e.g., 0.32~MB vs 1.01~MB for CESM, and 26.29~MB vs 36.64~MB for NYX), and consequently a higher Overall Compression Ratio (OCR). While floating edits occasionally yield a higher edit compression ratio due to the high redundancy of their explicit spatial indices, our quantized approach utilizes a dense bitmap representation, ultimately achieving a significantly smaller absolute compressed size. This reduction in edit size generally improves correction efficiency, as observed in CESM where the correction time decreases from 6.27~s to 3.83~s and in NYX from 710.89~s to 189.28~s. Consequently, the correction throughput is also improved in most cases, for example increasing from 7.88~MB/s to 12.90~MB/s on CESM and from 1.44~MB/s to 5.41~MB/s on NYX. However, these improvements are not uniform across all datasets, and in some cases the performance gain is limited or slightly reversed. This indicates that the effectiveness of quantization is also related to the topological complexity of the dataset. A more detailed analysis of this dependency is left for future work.}

\subsection{Time-Series Data Study}~\label{sec:time}
\begin{figure}[htb!]
    \centering
    \includegraphics[width=\linewidth]{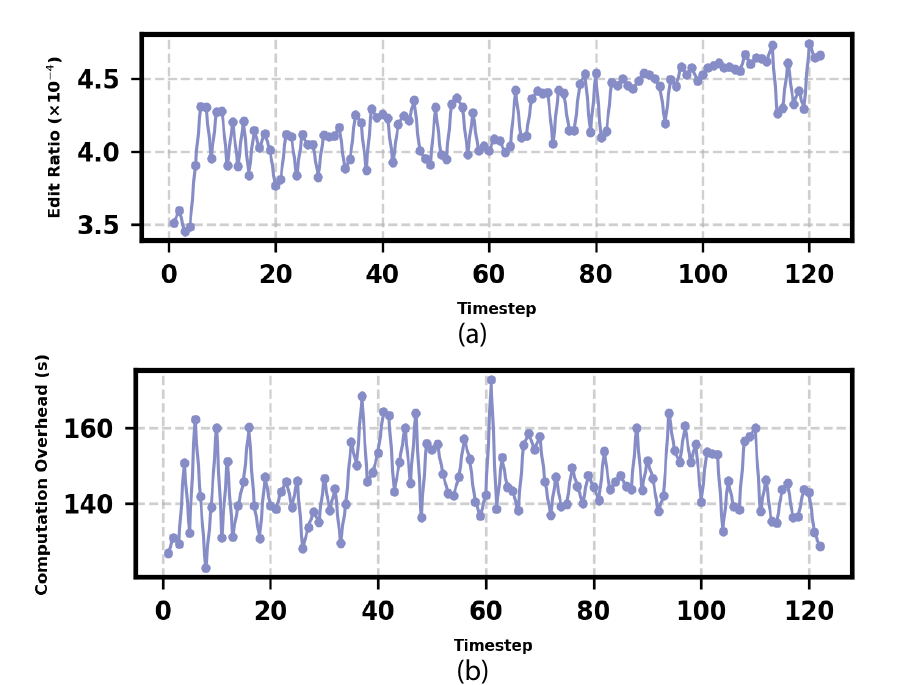}
    \caption{Processing overhead per time step for the turbulent combustion time series (122 timesteps, each of size $480\times720\times120$), using SZ3 as the base compressor with a relative error bound of $10^{-4}$. (a) Edit ratio required to preserve topological consistency across timesteps. (b) Computational overhead associated with maintaining topological consistency across timesteps.}
    \label{fig:time_series}
\end{figure}
We further evaluate DMTz on the turbulent combustion time-series dataset, which contains 122 scalar fields, each with a spatial resolution of $480\times720\times120$ stored in double precision and compressed by SZ3 with a relative error bound of $10^{-4}$.
Although each field corresponds to a single timestep and is relatively small in isolation, each requires an independent topological correction (as shown in Figure~\ref{fig:time_series}), making this dataset representative of large-scale temporal simulations.

In Figure~\ref{fig:time_series}, preserving MSC at each timestep requires editing a small percentage of data points; however, both the edit ratio (Figure~\ref{fig:time_series}(a)) and the corresponding computation overhead (Figure~\ref{fig:time_series}(b)) vary across timesteps and exhibit noticeable fluctuations. These variations arise because each timestep contains different data characteristics and topological configurations, resulting in different sensitivities to the same error bound. Consequently, the amount of required edits and the associated processing overhead jointly fluctuate over time.

\subsection{Performance Evaluation on GPUs} 
\begin{mdframed}
\noindent\textbf{Summary.} GPU acceleration delivers performance improvements for DMTz, with gradient pairing achieving up to 26× speedup, while saddle- and separatrix-related components exhibit lower acceleration (up to 9.69$\times$) due to their reliance on graph traversal and irregular memory access patterns.
\end{mdframed}

We assess the GPU performance of our algorithm on the NYX dataset by analyzing three major tasks:
(1) gradient pairing,
(2) false critical cell extraction and fixing, and
(3) false separatrix extraction and fixing.
Experiments were conducted on a single compute node equipped with four NVIDIA A100 GPUs on the Perlmutter supercomputer at the National Energy Research Scientific Computing Center (NERSC).
All components were implemented in C++ and CUDA, and we report averaged timings over 1,000 executions, with the data state reset between runs to eliminate initialization effects.
For comparison, we use an OpenMP-parallel implementation running with all 64 CPU threads on the same node as the baseline.

\begin{table}[htb!]
\centering
\caption{
Average timings, GPU acceleration, and compute/memory efficiency of different components of our method on the NYX dataset. All timing results are reported in seconds.
OpenMP results are measured using 64 CPU threads.
GPU Accel. is reported relative to OpenMP.
Efficiency is reported as compute efficiency (Comp. \%) and memory efficiency (Mem. \%).
}
\small
\renewcommand{\arraystretch}{1.2}
\setlength{\tabcolsep}{2pt}
\resizebox{\linewidth}{!}{
\begin{tabular}{ccccccc}
\hline
\textbf{Task} & \textbf{Sub-task} &
\textbf{OpenMP} &
\textbf{CUDA} &
\textbf{GPU Accel.} &
\textbf{Comp. \%} &
\textbf{Mem. \%} \\
\hline

\multirow{3}{*}{\begin{tabular}{c}Gradient\\Pairing\end{tabular}}
  & Vertices   & 0.770 & 0.042 & 18.33$\times$ & 86.95 & 31.50 \\
  & Edges      & 5.074 & 0.193 & 26.28$\times$ & 62.42 & 20.78 \\
  & Triangles  & 1.754 & 0.175 & 10.03$\times$ & 74.05 & 10.01 \\
\hline

\multirow{4}{*}{\begin{tabular}{c}False\\Critical Cells\\Extract + Fix\end{tabular}}
  & Minimum    & 0.035 & 0.002 & 17.54$\times$ & 39.49 & 48.03 \\
  & 1-Saddles  & 0.208 & 0.021 & 9.69$\times$  & 12.23 & 20.16 \\
  & 2-Saddles  & 0.251 & 0.090 & 2.78$\times$  & 8.08  & 8.81  \\
  & Maximum    & 0.095 & 0.040 & 2.36$\times$  & 11.92 & 13.02 \\
\hline

\multirow{3}{*}{\begin{tabular}{c}False\\Separatrices\\Extract + Fix\end{tabular}}
  & Descending     & 1.415 & 0.087 & 16.26$\times$ & 43.51 & 8.50 \\
  & Ascending      & 4.646 & 3.645 & 1.27$\times$  & 28.77 & 2.75 \\
  & Saddle-Saddle  & 16.98 & 7.884 & 2.15$\times$  & 38.29 & 4.67 \\
\hline
\end{tabular}}
\label{tab:accx}
\end{table}

Table~\ref{tab:accx} summarizes the performance of different tasks on the NYX dataset.
Gradient pairing achieves consistent and substantial acceleration over the 64-thread OpenMP baseline across vertices, edges, and triangles.
This improvement stems from the highly localized and regular computations involved in gradient pairing, which parallelize effectively on GPUs.

The extraction and fixing of false critical cells exhibit lower speedups, with performance varying across different critical cell types.
In particular, saddles show lower acceleration (up to 9.69$\times$) due to reduced parallel workload and less regular memory access patterns. Nevertheless, these tasks still benefit from GPU execution relative to the multi-threaded CPU baseline.
False separatrix extraction and fixing present the greatest performance challenges.
While descending separatrices achieve an acceleration of 16.26$\times$, ascending and saddle–saddle separatrix operations exhibit limited speedup up to 2.15$\times$.
This behavior is primarily due to their reliance on graph traversal and irregular memory access, which reduces GPU efficiency. Despite this, we retain separatrix-related computations on the GPU rather than offloading them to the CPU, as data transfer overheads would dominate the cost. 
Another observation is that the compute and memory efficiency vary across subtasks because of differences in workload size. 
A formal analysis of compute vs. memory-bound behavior remains challenging and is deferred to future work.

\begin{figure}
    \centering
    \includegraphics[width=\linewidth]{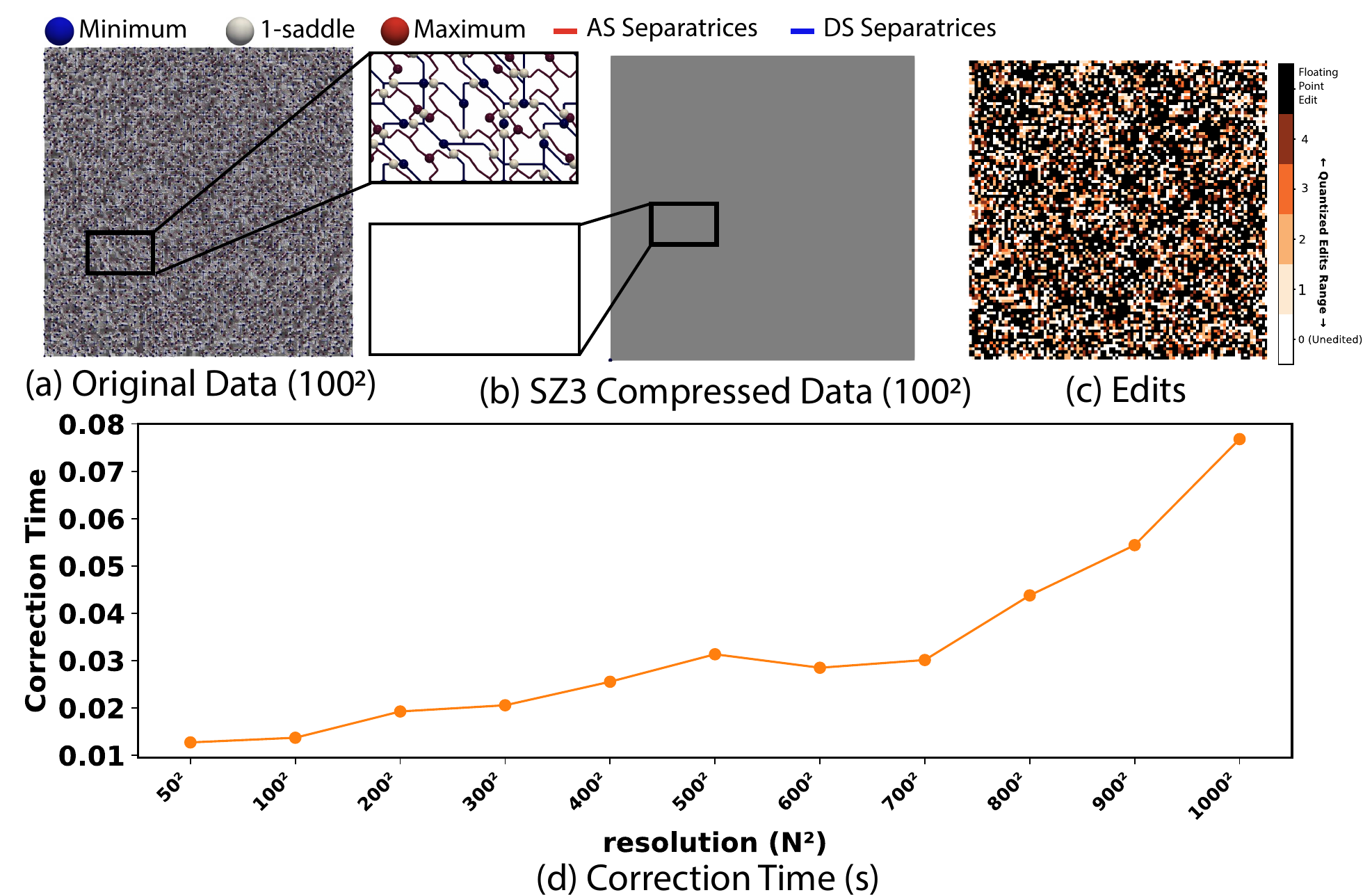}
    \caption{\updated{Evaluation of DMTz on an extreme synthetic case (2D random noise). Left: visual comparison of (a) original data, (b) SZ3-decompressed data ($100^2$), and (c) edits produced by DMTz. Right: (d) correction time across varying resolutions ($N^2$).}}
    \label{fig:worstcase}
\end{figure}

\subsection{Correction Behavior on Extreme Case}
\updated{Figure~\ref{fig:worstcase} evaluates an extreme synthetic case designed to stress the convergence behavior of DMTz. The input is a 2D random noise field whose scalar values are bounded within the prescribed error bound $\xi = 10^{-4}$. When compressed using SZ3 with the same error bound, the decompressed data becomes an almost completely flat field (Figure~\ref{fig:worstcase}(a, b)). In practice, users rarely select error bounds comparable to the data range, as such settings significantly distort local scalar ordering and suppress meaningful structures. Instead, this setting is used to stress the convergence behavior of DMTz when local scalar ordering is almost entirely destroyed.}

\updated{As shown in Figure~\ref{fig:worstcase}(d), the correction time increases with resolution, while the number of iterations grows moderately and stabilizes at larger sizes. This suggests that the runtime growth is primarily due to the increased data size processed per iteration rather than a proportional increase in iteration count. The edits produced by DMTz (Figure~\ref{fig:worstcase}(c)) are also widely distributed across the domain, indicating that a large portion of the field has been edited during the correction process. However, even in this extreme case, not all data points reach the maximum allowed number of edits per vertex (i.e., 6, indicated by the black regions in Figure~\ref{fig:worstcase}(c)), as discussed in Section~\ref{sec:convergence}. The observations also suggest that the convergence behavior is data-dependent, we leave a formal analysis of the convergence behavior for future work.}

\section{Limitations}
One limitation of our method is that it is tailored to the approach by Shivashankar et al.~\cite{SN12} for extracting MSC, thereby inheriting some of its drawbacks. For instance, their method can introduce many low-persistence critical points in noisy data, necessitating additional edits in our approach and leading to increased storage overhead. Additionally, the performance of false separatrix extraction remains a bottleneck, as it dominates the overall computation time in each iteration. Further optimization of this component is a direction for future work.

Our method may further improve the compression ratio by incorporating the simplification of MSC into the framework. By simplifying the target MSC (e.g., canceling low-persistence features before preservation), we can reduce the number of critical points and separatrices that need to be preserved, thereby reducing the number of edits and improving the compression ratio.

\section{Conclusions and Future Work}
We introduced DMTz, an iterative strategy for preserving key topological features of the Morse--Smale complex (MSC) in error-bounded lossy compression. We also designed a multitier preservation paradigm to support different applications' preservation needs. By enabling the multitiered preservation of extrema, saddles, and separatrices, our method provides flexible options for balancing compression efficiency and feature preservation.

We plan to improve our method in various aspects. First, we will focus on preserving simplified MSC under a specific persistence threshold. Second, we plan to optimize the extraction of false separatrices to improve overall performance. Third, we intend to better handle noisy data by integrating our workflow with the DMSC extraction method of Robins et al.~\cite{RRWS11}, reducing the impact of low-persistence critical points on computational and storage overhead.

\section*{Acknowledgment}

This work was supported by National Science Foundation grants OAC-2313122, OAC-2313123, OAC-2313124, and OAC-2442627. 
The work was also supported by the U.S. Department of
Energy, Office of Science, Advanced Scientific Computing
Research (ASCR), under contracts DE-SC0025677 and DE-SC0022753.  
This research used the resources of the National
Energy Research Scientific Computing Center (NERSC), a
Department of Energy User Facility using NERSC award
DDR-ERCAP 0034457.

\bibliographystyle{IEEEtran}
\bibliography{Refs-DMTz}

@inproceedings{sz,
  author={Liang, Xin and Di, Sheng and Tao, Dingwen and Li, Sihuan and Li, Shaomeng and Guo, Hanqi and Chen, Zizhong and Cappello, Franck},
  booktitle={Proceedings of 2018 IEEE International Conference on Big Data}, 
  
  title={Error-Controlled Lossy Compression Optimized for High Compression Ratios of Scientific Datasets}, 
  year={2018},
  volume={},
  number={},
  pages={438--447},
  doi={10.1109/BigData.2018.8622520}
}

@ARTICLE{sz3,
  author={Liang, Xin and Zhao, Kai and Di, Sheng and Li, Sihuan and Underwood, Robert and Gok, Ali M. and Tian, Jiannan and Deng, Junjing and Calhoun, Jon C. and Tao, Dingwen and Chen, Zizhong and Cappello, Franck},
  journal={IEEE Transactions on Big Data}, 
  title={{SZ3}: A Modular Framework for Composing Prediction-Based Error-Bounded Lossy Compressors},
  year={2023},
  volume={9},
  number={2},
  pages={485--498},
  doi={10.1109/TBDATA.2022.3201176}}

@inproceedings{Tao_2017,
  author={Tao, Dingwen and Di, Sheng and Chen, Zizhong and Cappello, Franck},
  booktitle={IPDPS '17: Proceedings of the 31st IEEE International Parallel and Distributed Processing Symposium}, 
  title={Significantly Improving Lossy Compression for Scientific Data Sets Based on Multidimensional Prediction and Error-Controlled Quantization}, 
  year={2017},
  pages={1129--1139},
  doi={10.1109/IPDPS.2017.115}
}

@inproceedings{Kai2021,
  author={Zhao, Kai and Di, Sheng and Dmitriev, Maxim and Tonellot, Thierry-Laurent D. and Chen, Zizhong and Cappello, Franck},
  booktitle={ICDE '21: Proceedings of the 37th IEEE International Conference on Data Engineering}, 
  title={Optimizing Error-Bounded Lossy Compression for Scientific Data by Dynamic Spline Interpolation}, 
  year={2021},
  volume={},
  number={},
  pages={1643--1654},
  doi={10.1109/ICDE51399.2021.00145}}

@inproceedings{Kai2020,
author = {Zhao, Kai and Di, Sheng and Liang, Xin and Li, Sihuan and Tao, Dingwen and Chen, Zizhong and Cappello, Franck},
title = {Significantly Improving Lossy Compression for {HPC} Datasets with Second-Order Prediction and Parameter Optimization},
year = {2020},
isbn = {9781450370523},
doi = {10.1145/3369583.3392688},
booktitle = {HPDC '20: Proceedings of the 29th International Symposium on High-Performance Parallel and Distributed Computing},
pages = {89–100},
numpages = {12},
location = {Stockholm, Sweden}
}

@article{zfp,
  author={Lindstrom, Peter},
  journal={IEEE Transactions on Visualization and Computer Graphics}, 
  title={Fixed-Rate Compressed Floating-Point Arrays}, 
  year={2014},
  volume={20},
  number={12},
  pages={2674--2683},
  doi={10.1109/TVCG.2014.2346458}
}

@ARTICLE{FPZIP,
  author={Lindstrom, Peter and Isenburg, Martin},
  journal={IEEE Transactions on Visualization and Computer Graphics}, 
  title={Fast and Efficient Compression of Floating-Point Data}, 
  year={2006},
  volume={12},
  number={5},
  pages={1245--1250},
  doi={10.1109/TVCG.2006.143}}

@ARTICLE{yan2023toposz,
  author={Yan, Lin and Liang, Xin and Guo, Hanqi and Wang, Bei},
  journal={IEEE Transactions on Visualization and Computer Graphics}, 
  title={{TopoSZ}: Preserving Topology in Error-Bounded Lossy Compression},
  year={2024},
  volume={30},
  number={1},
  pages={1302--1312},
  doi={10.1109/TVCG.2023.3326920}}

@inproceedings{li2026pmsz,
  author    = {Yuxiao Li and Mingze Xia and Xin Liang and {Bei Wang} and Robert Underwood and Sheng Di and Hemant Sharma and Dishant Beniwal and Franck Cappello and Hanqi Guo},
  title     = {{pMSz}: A Distributed Parallel Algorithm for Correcting {Morse-Smale} Segmentations for Lossy Compression},
  booktitle = {IPDPS '26: Proceedings of the 39th IEEE International Parallel and Distributed Processing Symposium},
  year      = {2026},
  note      = {To appear}
}

@inproceedings{tricoche2001continuous,
  title={Continuous topology simplification of planar vector fields},
  author={Tricoche, Xavier and Scheuermann, Gerik and Hagen, Hans},
  booktitle={VIS '01: Proceedings of IEEE Visualization},
  pages={159--166},
  year={2001},
  organization={IEEE}
}

@inproceedings{tricoche2000topology,
  title={A topology simplification method for {2D} vector fields},
  author={Tricoche, Xavier and Scheuermann, Gerik and Hagen, Hans},
  booktitle={VIS '00: Proceedings of IEEE Visualization},
  pages={359--366},
  year={2000},
  organization={IEEE}
}

@article{jacobson2012smooth,
  title={Smooth shape-aware functions with controlled extrema},
  author={Jacobson, Alec and Weinkauf, Tino and Sorkine, Olga},
  journal={Computer Graphics Forum},
  volume={31},
  number={5},
  pages={1577--1586},
  year={2012},
  organization={Wiley Online Library}
}

@article{theisel2002designing,
  title={Designing {2D} vector fields of arbitrary topology},
  author={Theisel, Holger},
  journal={Computer Graphics Forum},
  volume={21},
  number={3},
  pages={595--604},
  year={2002},
  organization={Wiley Online Library}
}

@ARTICLE{li_msz,
  author={Li, Yuxiao and Liang, Xin and Wang, Bei and Qiu, Yongfeng and Yan, Lin and Guo, Hanqi},
  journal={IEEE Transactions on Visualization and Computer Graphics}, 
  title={{MSz}: An Efficient Parallel Algorithm for Correcting {Morse-Smale} Segmentations in Error-Bounded Lossy Compressors}, 
  year={2025},
  volume={31},
  number={1},
  pages={130--140},
  doi={10.1109/TVCG.2024.3456337}}

@inproceedings{Liang_2020,
  author={Liang, Xin and Guo, Hanqi and Di, Sheng and Cappello, Franck and Raj, Mukund and Liu, Chunhui and Ono, Kenji and Chen, Zizhong and Peterka, Tom},
  booktitle={Proceedings of 2020 IEEE Pacific Visualization Symposium}, 
  
  title={Toward Feature-Preserving {2D} and {3D} Vector Field Compression}, 
  year={2020},
  volume={},
  number={},
  pages={81--90},
  doi={10.1109/PacificVis48177.2020.6431}
}

@inproceedings{msc,
author = {Edelsbrunner, Herbert and Harer, John and Natarajan, Vijay and Pascucci, Valerio},
title = {{Morse-Smale} Complexes for piecewise linear 3-manifolds},
year = {2003},
isbn = {1581136633},
doi = {10.1145/777792.777846},
booktitle = {SCG '03: Proceedings of the 19th Annual Symposium on Computational Geometry},
pages = {361--370},
numpages = {10},
}

@inproceedings{msc2,
  author = {Edelsbrunner, Herbert and Harer, John and Zomorodian, Afra},
  title = {Hierarchical {Morse} complexes for piecewise linear 2-manifolds},
  year = {2001},
  isbn = {158113357X},

  doi = {10.1145/378583.378626},
  booktitle = {SCG '01: Proceedings of the 17th Annual Symposium on Computational Geometry},
  pages = {70--79},
  numpages = {10}
}

@ARTICLE{Gyulassy_2007,
  author={Gyulassy, Attila and Duchaineau, Mark and Natarajan, Vijay and Pascucci, Valerio and Bringa, Eduardo and Higginbotham, Andrew and Hamann, Bernd},
  journal={IEEE Transactions on Visualization and Computer Graphics}, 
  title={Topologically Clean Distance Fields}, 
  year={2007},
  volume={13},
  number={6},
  pages={1432--1439},
  doi={10.1109/TVCG.2007.70603}}

@ARTICLE{Petruzza_2020,
  author={Petruzza, Steve and Gyulassy, Attila and Leventhal, Samuel and Baglino, John J. and Czabaj, Michael and Spear, Ashley D. and Pascucci, Valerio},
  journal={IEEE Transactions on Visualization and Computer Graphics}, 
  title={High-throughput feature extraction for measuring attributes of deforming open-cell foams}, 
  year={2020},
  volume={26},
  number={1},
  pages={140--150},
  doi={10.1109/TVCG.2019.2934620}}

@article{SHAHZAD2013,
author = {Shahzad, Faisal and Wittmann, Markus and Kreutzer, Moritz and Zeiser, Thomas and Hager, Georg and Wellein, Gerhard},
title = {A SURVEY OF CHECKPOINT/RESTART TECHNIQUES ON DISTRIBUTED MEMORY SYSTEMS},
journal = {Parallel Processing Letters},
volume = {23},
number = {04},
pages = {1340011},
year = {2013},
doi = {10.1142/S0129626413400112}
}

@article{Bhatia_2018,
author = {Bhatia, Harsh and Gyulassy, Attila G. and Lordi, Vincenzo and Pask, John E. and Pascucci, Valerio and Bremer, Peer-Timo},
title = {{TMS}: Comprehensive topological exploration for molecular and condensed-matter systems},
journal = {Journal of Computational Chemistry},
volume = {39},
number = {16},
pages = {936--952},
doi = {10.1002/jcc.25181},

year = {2018}
}

@ARTICLE{Shivashankar_2015,
  author={Shivashankar, Nithin and Pranav, Pratyush and Natarajan, Vijay and van de Weygaert, Rien and Bos, E. G. Patrick and Rieder, Steven},
  journal={IEEE Transactions on Visualization and Computer Graphics}, 
  title={Felix: A Topology Based Framework for Visual Exploration of Cosmic Filaments}, 
  year={2016},
  volume={22},
  number={6},
  pages={1745--1759},
  doi={10.1109/TVCG.2015.2452919}}

@ARTICLE{Günther_2014,
  author={Günther, David and Boto, Roberto A. and Contreras-Garcia, Juila and Piquemal, Jean-Philip and Tierny, Julien},
  journal={IEEE Transactions on Visualization and Computer Graphics}, 
  title={Characterizing Molecular Interactions in Chemical Systems}, 
  year={2014},
  volume={20},
  number={12},
  pages={2476--2485},
  doi={10.1109/TVCG.2014.2346403}}

@article{di2024survey,
author = {Di, Sheng and Liu, Jinyang and Zhao, Kai and Liang, Xin and Underwood, Robert and Zhang, Zhaorui and Shah, Milan and Huang, Yafan and Huang, Jiajun and Yu, Xiaodong and Ren, Congrong and Guo, Hanqi and Wilkins, Grant and Tao, Dingwen and Tian, Jiannan and Jin, Sian and Jian, Zizhe and Wang, Daoce and Rahman, Md Hasanur and Zhang, Boyuan and Song, Shihui and Calhoun, Jon and Li, Guanpeng and Yoshii, Kazutomo and Alharthi, Khalid and Cappello, Franck},
title = {A Survey on Error-Bounded Lossy Compression for Scientific Datasets},
year = {2025},
issue_date = {November 2025},
publisher = {Association for Computing Machinery},
address = {New York, NY, USA},
volume = {57},
number = {11},
issn = {0360-0300},
doi = {10.1145/3733104},
journal = {ACM Comput. Surv.},
pages = {1--38, article. no. 287}
}

@misc{zstd,
  key = {zstd},
  title = {{ZSTD}},
  howpublished = {\url{https://github.com/facebook/zstd/releases}},
  note = {Accessed: January 2026}
}

@inproceedings{ISABELA,
author = {Lakshminarasimhan, Sriram and Shah, Neil and Ethier, Stephane and Klasky, Scott and Latham, Rob and Ross, Rob and Samatova, Nagiza F.},
title = {Compressing the incompressible with {ISABELA}: In-situ reduction of spatio-temporal data},
year = {2011},
isbn = {9783642233999},
booktitle = {Euro-Par '11: Proceedings of the 17th International Conference on Parallel Processing - Volume Part I},
pages = {366--379},
numpages = {14},
}

@article{MGARD,
   title={{MGARD}: A multigrid framework for high-performance, error-controlled data compression and refactoring},
   volume={24},
   ISSN={2352-7110},
   DOI={10.1016/j.softx.2023.101590},
   journal={SoftwareX},
   publisher={Elsevier BV},
   author={Gong, Qian and Chen, Jieyang and Whitney, Ben and Liang, Xin and Reshniak, Viktor and Banerjee, Tania and Lee, Jaemoon and Rangarajan, Anand and Wan, Lipeng and Vidal, Nicolas and Liu, Qing and Gainaru, Ana and Podhorszki, Norbert and Archibald, Richard and Ranka, Sanjay and Klasky, Scott},
   year={2023},
   pages={101590} }

@INPROCEEDINGS {soler2018topologically,
author = {M. Soler and M. Plainchault and B. Conche and J. Tierny},
booktitle = {Proceedings of 2018 IEEE Pacific Visualization Symposium},
title = {Topologically Controlled Lossy Compression},
year = {2018},
volume = {},
pages = {46--55},
doi = {10.1109/PacificVis.2018.00015}
}

@ARTICLE{LiangDCRLOCPG23,
  author={Liang, Xin and Di, Sheng and Cappello, Franck and Raj, Mukund and Liu, Chunhui and Ono, Kenji and Chen, Zizhong and Peterka, Tom and Guo, Hanqi},
  journal={IEEE Transactions on Visualization and Computer Graphics}, 
  title={Toward Feature-Preserving Vector Field Compression}, 
  year={2023},
  volume={29},
  number={12},
  pages={5434--5450},
  doi={10.1109/TVCG.2022.3214821}}

@article{cp,
author = {Thomas Banchoff},
title = {Critical Points and Curvature for Embedded Polyhedra},
volume = {1},
journal = {Journal of Differential Geometry},
number = {3--4},
publisher = {Lehigh University},
pages = {245--256},
year = {1967},
doi = {10.4310/jdg/1214428092}
}

@article{FORMAN199890,
title = {Morse Theory for Cell Complexes},
journal = {Advances in Mathematics},
volume = {134},
number = {1},
pages = {90--145},
year = {1998},
issn = {0001-8708},
doi = {10.1006/aima.1997.1650},
author = {Robin Forman}
}

@ARTICLE{RRWS11,
  author={Robins, Vanessa and Wood, Peter John and Sheppard, Adrian P.},
  journal={IEEE Transactions on Pattern Analysis and Machine Intelligence}, 
  title={Theory and Algorithms for Constructing Discrete {Morse} Complexes from Grayscale Digital Images}, 
  year={2011},
  volume={33},
  number={8},
  pages={1646-1658},
  doi={10.1109/TPAMI.2011.95}}

@article{SN12,
author = {Shivashankar, Nithin and Natarajan, Vijay},
title = {Parallel Computation of {3D} {Morse-Smale} Complexes},
journal = {Computer Graphics Forum},
volume = {31},
number = {3pt1},
pages = {965--974},
doi = {10.1111/j.1467-8659.2012.03089.x},

year = {2012}
}

@article{Lan_ivt,
author = {Lan, Fangfei and Gamelin, Brandi and Yan, Lin and Wang, Jiali and Wang, Bei and Guo, Hanqi},
title = {Topological Characterization and Uncertainty Visualization of Atmospheric Rivers},
journal = {Computer Graphics Forum},
volume = {43},
number = {3},
pages = {e15084},

doi = {10.1111/cgf.15084},
year = {2024}
}

@ARTICLE{Dora_2013,
  author={Doraiswamy, Harish and Natarajan, Vijay and Nanjundiah, Ravi S.},
  journal={IEEE Transactions on Visualization and Computer Graphics}, 
  title={An Exploration Framework to Identify and Track Movement of Cloud Systems}, 
  year={2013},
  volume={19},
  number={12},
  pages={2896-2905},
  doi={10.1109/TVCG.2013.131}}

@ARTICLE{MSS,
  author={Maack, Robin G. C. and Lukasczyk, Jonas and Tierny, Julien and Hagen, Hans and Maciejewski, Ross and Garth, Christoph},
  journal={IEEE Transactions on Visualization and Computer Graphics}, 
  title={Parallel Computation of Piecewise Linear {Morse-Smale} Segmentations}, 
  year={2024},
  volume={30},
  number={4},
  pages={1942--1955},
  doi={10.1109/TVCG.2023.3261981}}

@article{Forman_dmt,
author = {Forman, Robin},
year = {2002},
pages = {B48c, 35p.},
title = {A User's Guide To Discrete {Morse} Theory},
volume = {48},
journal = {Sém. Lothar. Combin.}
}

@ARTICLE{Bremer_combustion,
  author={Bremer, Peer-Timo and Weber, Gunther and Tierny, Julien and Pascucci, Valerio and Day, Marc and Bell, John},
  journal={IEEE Transactions on Visualization and Computer Graphics}, 
  title={Interactive Exploration and Analysis of Large-Scale Simulations Using Topology-Based Data Segmentation}, 
  year={2011},
  volume={17},
  number={9},
  pages={1307--1324},
  
  doi={10.1109/TVCG.2010.253}}

@ARTICLE{Bremer_flames,
  author={Bremer, Peer-Timo and Weber, Gunther and Pascucci, Valerio and Day, Marc and Bell, John},
  journal={IEEE Transactions on Visualization and Computer Graphics}, 
  title={Analyzing and Tracking Burning Structures in Lean Premixed Hydrogen Flames}, 
  year={2010},
  volume={16},
  number={2},
  pages={248--260},
  doi={10.1109/TVCG.2009.69}}

@inproceedings{Edelsbrunner_persistence,
  author={Edelsbrunner, H. and Letscher, D. and Zomorodian, A.},
  booktitle={Proceedings of 41st Annual Symposium on Foundations of Computer Science}, 
  title={Topological persistence and simplification}, 
  year={2000},
  volume={},
  number={},
  pages={454-463},
  doi={10.1109/SFCS.2000.892133}}

@ARTICLE{pont2023,
  author={Pont, Mathieu and Tierny, Julien},
  journal={IEEE Transactions on Visualization and Computer Graphics}, 
  title={Wasserstein Auto-Encoders of Merge Trees (and Persistence Diagrams)}, 
  year={2024},
  volume={30},
  number={9},
  pages={6390-6406},
  doi={10.1109/TVCG.2023.3334755}}

@article{De_2015,
author = {De Floriani, Leila and Fugacci, Ulderico and Iuricich, Federico and Magillo, Paola},
title = {{Morse} complexes for shape segmentation and homological analysis: discrete models and algorithms},
journal = {Computer Graphics Forum},
volume = {34},
number = {2},
pages = {761--785},
doi = {10.1111/cgf.12596},
year = {2015}
}

@inproceedings{liu_2021,
  author={Liu, Jinyang and Di, Sheng and Zhao, Kai and Jin, Sian and Tao, Dingwen and Liang, Xin and Chen, Zizhong and Cappello, Franck},
  booktitle={Proceedings of 2021 IEEE International Conference on Cluster Computing (CLUSTER)}, 
  title={Exploring Autoencoder-based Error-bounded Compression for Scientific Data}, 
  year={2021},
  volume={},
  number={},
  pages={294-306},
  doi={10.1109/Cluster48925.2021.00034}}

@inproceedings{theisel2003combining,
  author={Theisel, H. and Rossl, C. and Seidel, H.-P.},
  booktitle={Proceedings of the 11th Pacific Conference on Computer Graphics and Applications}, 
  title={Combining topological simplification and topology preserving compression for {2D} vector fields}, 
  year={2003},
  volume={},
  number={},
  pages={419--423},
  doi={10.1109/PCCGA.2003.1238287}}

@inproceedings{Weinkauf_05,
  author={Weinkauf, T. and Theisel, H. and Shi, K. and Hege, H.-C. and Seidel, H.-P.},
  booktitle={VIS '05: Proceedings of the 16th IEEE Visualization Conference}, 
  title={Extracting higher order critical points and topological simplification of {3D} vector fields}, 
  year={2005},
  pages={559--566},
  doi={10.1109/VISUAL.2005.1532842}}

@article{Weinkauf_10,
author = {Weinkauf, Tino and Gingold, Yotam and Sorkine, Olga},
title = {Topology-based Smoothing of {2D} Scalar Fields with {C1}-Continuity},
journal = {Computer Graphics Forum},
volume = {29},
number = {3},
pages = {1221--1230},
doi = {10.1111/j.1467-8659.2009.01702.x},
year = {2010}
}

@ARTICLE{Gunther_14,
  author={Günther, David and Jacobson, Alec and Reininghaus, Jan and Seidel, Hans-Peter and Sorkine-Hornung, Olga and Weinkauf, Tino},
  journal={IEEE Transactions on Visualization and Computer Graphics}, 
  title={Fast and Memory-Efficient Topological Denoising of {2D} and {3D} Scalar Fields}, 
  year={2014},
  volume={20},
  number={12},
  pages={2585--2594},
  doi={10.1109/TVCG.2014.2346432}}

@ARTICLE{Tierny_12,
  author={Tierny, Julien and Pascucci, Valerio},
  journal={IEEE Transactions on Visualization and Computer Graphics}, 
  title={Generalized Topological Simplification of Scalar Fields on Surfaces}, 
  year={2012},
  volume={18},
  number={12},
  pages={2005--2013},
  doi={10.1109/TVCG.2012.228}}

@ARTICLE{Gyulassy_06,
  author={Gyulassy, A. and Vijay Natarajan and Pascucci, V. and Bremer, P.-T. and Hamann, B.},
  journal={IEEE Transactions on Visualization and Computer Graphics}, 
  title={A topological approach to simplification of three-dimensional scalar functions}, 
  year={2006},
  volume={12},
  number={4},
  pages={474--484},
  
  doi={10.1109/TVCG.2006.57}}

@article{masood2021overview,
  title={An overview of the topology toolkit},
  author={Bin Masood, Talha and Budin, Joseph and Falk, Martin and Favelier, Guillaume and Garth, Christoph and Gueunet, Charles and Guillou, Pierre and Hofmann, Lutz and Hristov, Petar and Kamakshidasan, Adhitya and Kappe, Christopher and Klacansky, Pavol and Laurin, Patrick and Levine, Joshua A and Lukasczyk, Jonas and Sakurai, Daisuke and Soler, Maxime and Steneteg, Peter and Tierny, Julien and Usher, Will and Vidal, Jules and Wozniak, Michal},
  journal={Topological Methods in Data Analysis and Visualization VI: Theory, Applications, and Software},
  pages={327--342},
  year={2021}
}

@ARTICLE{Kissi_25,
  author={Kissi, Mohamed and Pont, Mathieu and Levine, Joshua A. and Tierny, Julien},
  journal={IEEE Transactions on Visualization and Computer Graphics}, 
  title={A Practical Solver for Scalar Data Topological Simplification}, 
  year={2025},
  volume={31},
  number={1},
  pages={97--107},
  doi={10.1109/TVCG.2024.3456345}}

@ARTICLE{gorski2025general,
author={Gorski, Nathaniel and Liang, Xin and Guo, Hanqi and Yan, Lin and Wang, Bei},
journal={ IEEE Transactions on Visualization and  Computer Graphics },
title={A General Framework for Augmenting Lossy Compressors With Topological Guarantees},
year={2025},
volume={31},
number={06},
ISSN={1941-0506},
pages={3693--3705},
doi={10.1109/TVCG.2025.3567054},
publisher={IEEE Computer Society},
address={Los Alamitos, CA, USA}}

@article{CARR200375,
title = {Computing contour trees in all dimensions},
journal = {Computational Geometry},
volume = {24},
number = {2},
pages = {75--94},
year = {2003},
issn = {0925-7721},
doi = {10.1016/S0925-7721(02)00093-7},
author = {Hamish Carr and Jack Snoeyink and Ulrike Axen}
}

@article{NeRVI,
title = {{NeRVI}: Compressive neural representation of visualization images for communicating volume visualization results},
journal = {Computers \& Graphics},
volume = {116},
pages = {216--227},
year = {2023},
issn = {0097-8493},
doi = {10.1016/j.cag.2023.08.024},
author = {Pengfei Gu and Danny Z. Chen and Chaoli Wang}
}

@article{Neurcomp,
author = {Lu, Y. and Jiang, K. and Levine, J. A. and Berger, M.},
title = {Compressive Neural Representations of Volumetric Scalar Fields},
journal = {Computer Graphics Forum},
volume = {40},
number = {3},
pages = {135--146},
doi = {10.1111/cgf.14295},

year = {2021}
}

@ARTICLE{Liu_AE,
  author={Liu, Tong and Wang, Jinzhen and Liu, Qing and Alibhai, Shakeel and Lu, Tao and He, Xubin},
  journal={IEEE Transactions on Big Data}, 
  title={High-Ratio Lossy Compression: Exploring the Autoencoder to Compress Scientific Data}, 
  year={2023},
  volume={9},
  number={1},
  pages={22--36},
  doi={10.1109/TBDATA.2021.3066151}}

@inproceedings{Qoz,
  author={Liu, Jinyang and Di, Sheng and Zhao, Kai and Liang, Xin and Chen, Zizhong and Cappello, Franck},
  booktitle={SC '22: Proceedings of International Conference for High Performance Computing, Networking, Storage and Analysis}, 
  title={Dynamic Quality Metric Oriented Error Bounded Lossy Compression for Scientific Datasets}, 
  year={2022},
  volume={},
  number={},
  pages={62:1--62:15},
  doi={10.1109/SC41404.2022.00067}}

@INPROCEEDINGS {liu2023srnsz,
author = { Liu, Jinyang and Di, Sheng and Jin, Sian and Zhao, Kai and Liang, Xin and Chen, Zizhong and Cappello, Franck },
booktitle = {Proceedings of 2023 IEEE International Conference on Big Data},
title = {Scientific Error-bounded Lossy Compression with Super-resolution Neural Networks},
year = {2023},
pages = {229-236},
doi = {10.1109/BigData59044.2023.10386682}
}

@ARTICLE{TTHRESH,
  author={Ballester-Ripoll, Rafael and Lindstrom, Peter and Pajarola, Renato},
  journal={IEEE Transactions on Visualization and Computer Graphics}, 
  title={{TTHRESH}: Tensor Compression for Multidimensional Visual Data}, 
  year={2020},
  volume={26},
  number={9},
  pages={2891--2903},
  doi={10.1109/TVCG.2019.2904063}}

@inproceedings{Xia24,
  author={Xia, Mingze and Di, Sheng and Cappello, Franck and Jiao, Pu and Zhao, Kai and Liu, Jinyang and Wu, Xuan and Liang, Xin and Guo, Hanqi},
  booktitle={ICDE '24: Proceedings of the 40th IEEE International Conference on Data Engineering}, 
  title={Preserving Topological Feature with Sign-of-Determinant Predicates in Lossy Compression: A Case Study of Vector Field Critical Points}, 
  year={2024},

  pages={4979--4992},
  doi={10.1109/ICDE60146.2024.00378}
}

@INPROCEEDINGS {Xia25,
author = { Xia, Mingze and Wang, Bei and Li, Yuxiao and Jiao, Pu and Liang, Xin and Guo, Hanqi },
booktitle = {ICDE '25: Proceedings of the 41st IEEE International Conference on Data Engineering},
title = {{TspSZ}: An Efficient Parallel Error-Bounded Lossy Compressor for Topological Skeleton Preservation},
year = {2025},
volume = {},
ISSN = {},
pages = {3682--3695},
doi = {10.1109/ICDE65448.2025.00275},
}

@inproceedings{huang2023cuszp,
author = {Huang, Yafan and Di, Sheng and Yu, Xiaodong and Li, Guanpeng and Cappello, Franck},
title = {{cuSZp}: An Ultra-fast GPU Error-bounded Lossy Compression Framework with Optimized End-to-End Performance},
year = {2023},
isbn = {9798400701092},
doi = {10.1145/3581784.3607048},
booktitle = {SC '23: Proceedings of the International Conference for High Performance Computing, Networking, Storage and Analysis},
articleno = {43},
numpages = {13},
pages={43:1--43:13},
}

@inproceedings{huang2024cuszp2,
author = {Huang, Yafan and Di, Sheng and Li, Guanpeng and Cappello, Franck},
title = {{cuSZp2}: A GPU Lossy Compressor with Extreme Throughput and Optimized Compression Ratio},
year = {2024},
isbn = {9798350352917},
doi = {10.1109/SC41406.2024.00021},
booktitle = {SC '24: Proceedings of the International Conference for High Performance Computing, Networking, Storage, and Analysis},
articleno = {15},
numpages = {18},
pages={15:1--15:18}
}

@inproceedings{SPERR,
  author={Li, Shaomeng and Lindstrom, Peter and Clyne, John},
  booktitle={IPDPS '23: Proceedings of the 37th IEEE International Parallel and Distributed Processing Symposium}, 
  
  title={Lossy Scientific Data Compression With {SPERR}}, 
  year={2023},
  pages={1007--1017},
  doi={10.1109/IPDPS54959.2023.00104}}

@ARTICLE{Wang_Robustness,
  author={Skraba, Primoz and Wang, Bei and Chen, Guoning and Rosen, Paul},
  journal={IEEE Transactions on Visualization and Computer Graphics}, 
  title={Robustness-Based Simplification of {2D} Steady and Unsteady Vector Fields}, 
  year={2015},
  volume={21},
  number={8},
  pages={930--944},
  doi={10.1109/TVCG.2015.2440250}}

@article{XU201438,
title = {Medical image fusion using multi-level local extrema},
journal = {Information Fusion},
volume = {19},
pages = {38--48},
year = {2014},
issn = {1566-2535},
doi = {10.1016/j.inffus.2013.01.001},
author = {Zhiping Xu}
}

@book{Edelsbrunner_10,
author = {Edelsbrunner, Herbert and Harer, John},
year = {2009},
publisher = {American Mathematical Society},
title = {Computational Topology: An Introduction},
isbn = {978-0-8218-4925-5},
doi = {10.1007/978-3-540-33259-6_7}
}

@inproceedings{huang2025lscomp,
author = {Huang, Yafan and Di, Sheng and Underwood, Robert and Myint, Peco and Chu, Miaoqi and Li, Guanpeng and Schwarz, Nicholas and Cappello, Franck},
title = {{lsCOMP}: Efficient Light Source Compression},
year = {2025},
isbn = {9798400714665},
doi = {10.1145/3712285.3759814},
booktitle = {SC '25: Proceedings of the International Conference for High Performance Computing, Networking, Storage and Analysis},
pages = {2006--2023},
numpages = {18}
}

\begin{IEEEbiography}[{\includegraphics[width=1in,height=1.25in,clip,keepaspectratio]{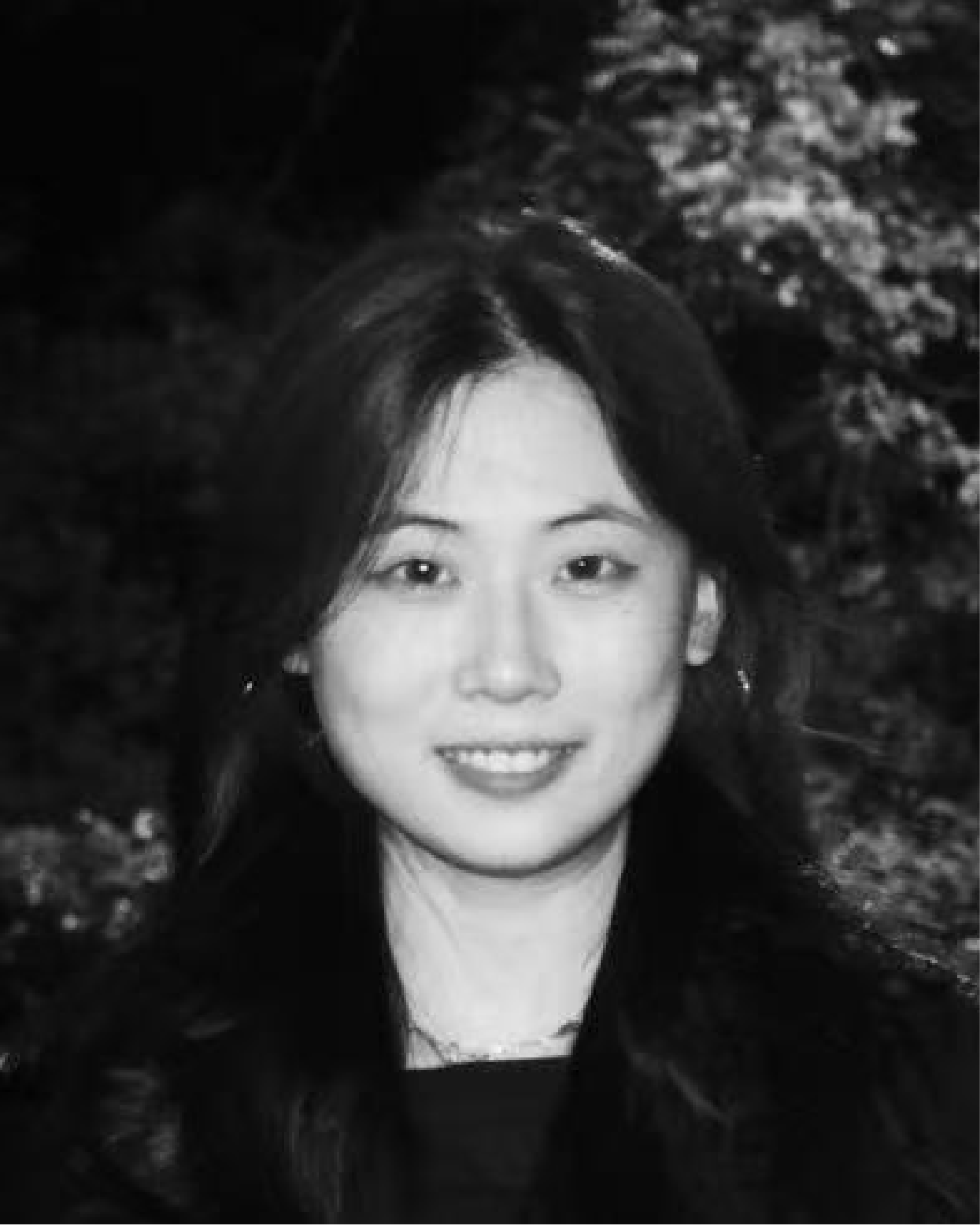}}]{Yuxiao Li}(Student Member, IEEE) received the bachelor’s degree from Sichuan University in 2021, and the master’s degree from Fudan University in 2023. She is currently pursuing the Ph.D. degree at The Ohio State University. Her research interests include data compression, parallel computing, and scientific visualization.
\end{IEEEbiography}

\begin{IEEEbiography}[{\includegraphics[width=1in,height=1.25in,clip,keepaspectratio]{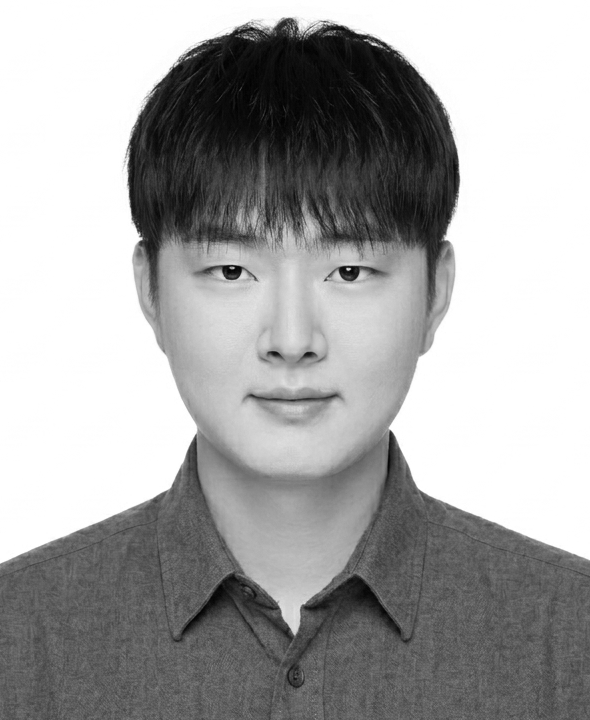}}]{Mingze Xia}(Student Member, IEEE) received the B.Math. degree in statistics from the University of Waterloo, Canada, in 2017, and the master’s degree in data science from The University of Melbourne, Australia, in 2019. From 2020 to 2022, he was a Research Fellow with Duke Kunshan University and worked as a Data Scientist in industry. He is currently pursuing the Ph.D. degree at Oregon State University. His research interests include data compression, parallel computing, and scientific visualization.
\end{IEEEbiography}

\begin{IEEEbiography}[{\includegraphics[width=1in,height=1.25in,clip,keepaspectratio]{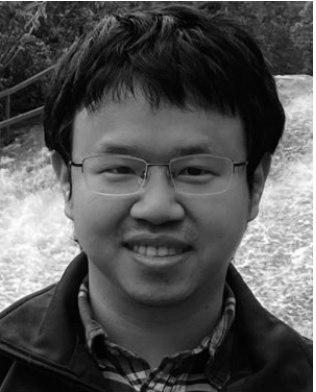}}]{Xin Liang} (Member, IEEE) received the bachelor’s degree from Peking University in 2014 and the Ph.D. degree from the University of California, Riverside, in 2019. He is an Associate Professor with the School of Electrical Engineering and Computer Science at Oregon State University. Prior to that, he worked as an Assistant Professor at the University of Kentucky and the Missouri University of Science and Technology, and as a Computer/Data Scientist in the Workflow Systems Group at Oak Ridge National Laboratory (ORNL). His research interests include high-performance computing, parallel and distributed systems, scientific data management and reduction, big data analytics, and scientific visualization. He has received two Best Paper Awards from IEEE Cluster, a Best Paper Finalist recognition from ACM ICS, a Best Paper Award from IEEE IPDPS, and a Best Paper Award from IEEE Transactions on Big Data (TBD). He has also received the NSF CRII Award, the NSF EPSCoR Fellowship, the NSF CAREER Award, and the IEEE Computer Society TCHPC Early Career Researchers Award for Excellence in High Performance Computing in 2024.
\end{IEEEbiography}

\begin{IEEEbiography}[{\includegraphics[width=1in,height=1.25in,clip,keepaspectratio]{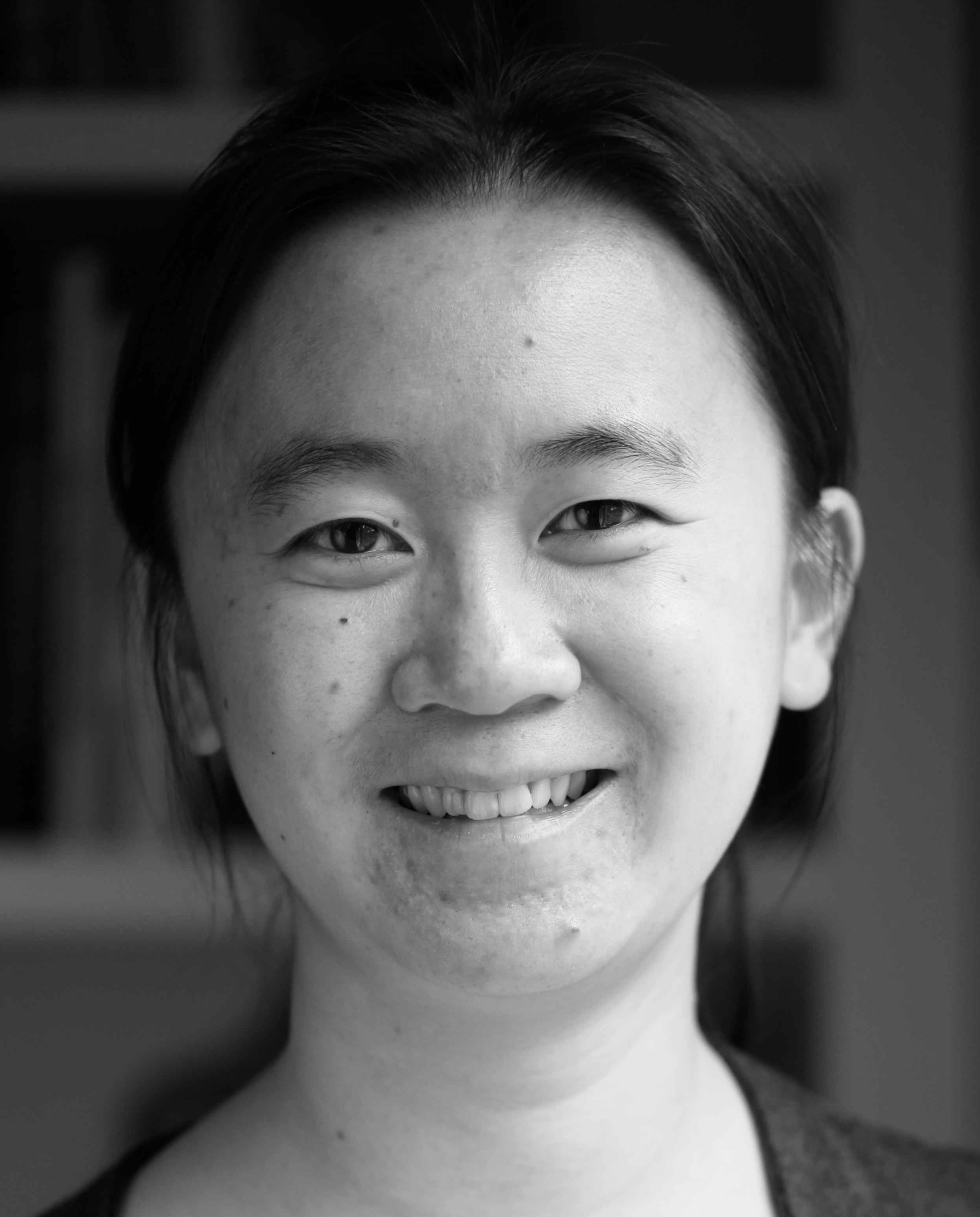}}]{Bei Wang} (Member, IEEE) received the Ph.D. degree in computer science from Duke University. She is an Associate Professor in the School of Computing, an Adjunct Associate Professor in the Department of Mathematics, and a faculty member of the Scientific Computing and Imaging (SCI) Institute at the University of Utah. Her research lies at the intersection of topological data analysis, data visualization, and computational topology, with a focus on integrating topological, geometric, statistical, data mining, and machine learning methods with visualization to enable scientific discovery in large and complex datasets. Her work has been supported by multiple awards from the NSF, NIH, and DOE. She received a DOE Early Career Research Program Award in 2020, an NSF CAREER Award in 2022, and the Presidential Early Career Award for Scientists and Engineers (PECASE) from President Biden in 2024, the highest honor bestowed by the U.S. government on early-career scientists and engineers.
\end{IEEEbiography}

\begin{IEEEbiography}[{\includegraphics[width=1in,height=1.25in,clip,keepaspectratio]{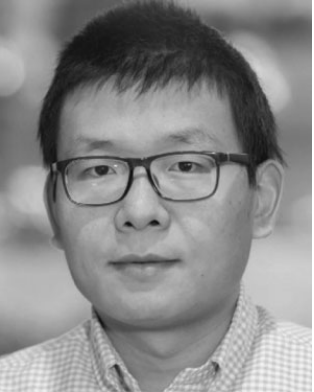}}]{Hanqi Guo} (Member, IEEE) received the B.S. degree in mathematics and applied mathematics from Beijing University of Posts and Telecommunications in 2009, and the Ph.D. degree in computer science from Peking University in 2014. He is an Associate Professor with the Department of Computer Science and Engineering at The Ohio State University. His research interests include data analysis, visualization, and machine learning for scientific data. He was a recipient of the DOE Early Career Research Program (ECRP) Award in 2022 and has received multiple best paper awards at premier visualization conferences.
\end{IEEEbiography}

\vspace{11pt}

\vfill

\end{document}